\documentclass[]{aa}
\usepackage{graphicx}
\usepackage{txfonts}
\usepackage{longtable}
\usepackage{lscape}
\usepackage{natbib}
\usepackage{lscape}
\usepackage{amsmath}
\usepackage{wasysym}
\usepackage{graphics}
\usepackage{txfonts}
\usepackage{color}
\usepackage{times}
\usepackage{parskip}
\usepackage{pdflscape}
\usepackage{geometry}
\usepackage{marginnote}
\usepackage[table,dvipsnames]{xcolor}
\usepackage{stfloats}
\usepackage{multicol}
\usepackage{floatrow}
\usepackage{etoolbox}
\usepackage{caption}
\usepackage[title]{appendix}
\usepackage{tabularx}
\usepackage[linesnumbered,ruled,vlined]{algorithm2e}
\usepackage{amsfonts}
\usepackage{listings}
\newenvironment{Figure1}
  {\par\medskip\noindent\minipage{\linewidth}}
  {\endminipage\par\medskip}

\newfont{\nlx}{cmssdc10 scaled 900}
\newfont{\mlx}{cmssdc10 scaled 1032}
\newfont{\mfont}{cmssdc10 scaled 760}
\definecolor{myblue1}{rgb}{0.0,0.604,0.831} 
\definecolor{myblue2}{rgb}{0.0,0.49,0.6745}
\definecolor{myblue3}{rgb}{0.0156,0.4078,0.9921}
\definecolor{myblue4}{rgb}{0.0,0.44,0.87}
\definecolor{myred1}{rgb}{0.529,0.019,0.017}
\definecolor{mycyan}{rgb}{0.63921569,0.0,0.48235294}

\newcommand{\brem}[1]{\textcolor{black}{\nlx #1}}

\def\D4000{$D_{4000}$}

\def\rr{R$^{\star}$}

\newcommand{\sbb}{mag/$\sq\arcsec$}

\def\sstar{$\Sigma_{\star,\mathrm{B}}$}

\newfont{\hvss}{cmssdc10 scaled 1540}
\def\?{{\bf\color{red}?}}
\def\reff{R$_{\rm eff}$}

\def\mmu{$\langle \mathrm{I}_{i} \rangle$}

\def\lrmax{$\rm L(R^{\star}_{\rm max})$}
\def\irr{$\rm I(R^{\star})$}
\def\meff{$\mu_{\rm eff}$}
\def\SL{\nlx SL\rm}
\def\iSP{\nlx iSP\rm}
\def\iSPs{\nlx iSPs\rm}
\def\ifit{i{\sc Fit}}
\def\eSP{\nlx eSP\rm}
\def\tspack{{\sc TSPACK}}
\def\minpack{{\sc MINPACK}}
\def\galfit{{\sc GALFIT}}
\def\mlr{$\cal{M}/\cal{L}$}
\def\snr{S/N}
\def\mulim{$\mu_{\rm lim}$}
\def\murlim{$\rm \mu(R^{\star}_{max})$}
\makeatletter
\preto{\@verbatim}{\topsep=0pt \partopsep=0pt }
\makeatother

\begin{document} 

\title{A new fitting concept for the robust determination of S\'ersic model parameters}
   \author{Iris Breda
          \inst{1,2}
          \and
          Polychronis Papaderos
          \inst{1,3,4,5}
          \and
          Jean Michel Gomes 
          \inst{1} 
          \and
          Stergios Amarantidis 
          \inst{4,5} 
          }
   \institute{Instituto de Astrof\'{i}sica e Ci\^{e}ncias do Espaço - Centro de Astrof\'isica da Universidade do Porto, Rua das Estrelas, 4150-762 Porto, Portugal
         \and
   Departamento de F\'isica e Astronomia, Faculdade de Ci\^encias, Universidade do Porto, Rua do Campo Alegre, 4169-007 Porto, Portugal
        \and 
   Guest professor, University of Vienna, Department of Astrophysics, T\"urkenschanzstr. 17, 1180 Vienna, Austria
         \and
   Instituto de Astrofísica e Ciências do Espaço, Universidade de Lisboa, OAL, Tapada da Ajuda, PT1349-018 Lisboa, Portugal
         \and 
   Departamento de Física, Faculdade de Ciências da Universidade de Lisboa, Edifício C8, Campo Grande, PT1749-016 Lisboa, Portugal         
   \\            
             \email{\tiny iris.breda@astro.up.pt,papaderos@astro.up.pt,jean@astro.up.pt,samarant@oal.ul.pt}
             }


\abstract
{The S\'{e}rsic law (\SL) offers a versatile, widely used functional form for the structural 
characterization of galaxies near and far. 
Whereas fitting this three-parameter function to galaxies with a genuine \SL\ luminosity distribution 
(e.g., several local early-type galaxies--ETGs) yields a robust determination of the 
S\'ersic exponent $\eta$ and effective surface brightness $\mu_{\rm eff}$, this is not necessarily 
the case for galaxies whose surface brightness profiles (SBPs) appreciably deviate, either in their centers or over an extended radius interval, from the \SL\ (e.g., ETGs with a ``depleted'' core and nucleated dwarf ellipticals, or most late-type galaxies-LTGs). 
In this general case of ``imperfect'' \SL\ profiles, the best-fitting solution may significantly depend on the radius (or surface brightness) interval fit, the photometric passbands considered and the specifics of the fitting procedure 
(photometric uncertainties of SBP data points or image pixels, and corrections for point spread function (PSF) 
convolution effects). Such uncertainties may then affect, in a non-easily predictable manner, automated structural 
studies of large heterogeneous galaxy samples and introduce a scatter, if not a bias, in galaxy scaling relations 
and their evolution across redshift ($z$).}
{Our goal is to devise a fitting concept that permits a robust determination of the equivalent \SL\ model 
for the general case of galaxies with imperfect \SL\ profiles.}
{The distinctive feature of the concept proposed here (\ifit) is that the fit    
is not constrained through standard $\chi^2$ minimization between an observed SBP and the \SL\ model 
of it, but instead through the search for the best match between the observationally determined and theoretically expected radial variation of the mean surface brightness and light growth curve.
This approach ensures quick convergence to a unique solution for both perfect and imperfect 
S\'ersic profiles, even shallow and resolution-degraded SBPs. \ifit\ allows for correction of PSF convolution
effects, offering the user the option of choosing between a Moffat, Gaussian, or user-supplied PSF.
\ifit, which is a standalone FORTRAN code, can be applied to any SBP that is provided in ASCII format and it has the capability of convenient graphical storage of its output. The \ifit\ distribution package is supplemented with an auxiliary SBP derivation tool in python.}
{\ifit\ has been extensively tested on synthetic data with a S\'{e}rsic index $0.3 \leq \eta \leq 4.2$ and an effective radius $1 \leq \mathrm{R}_{\rm eff}\,\,(\arcsec) \leq 20$. Applied to non PSF-convolved data, \ifit\ can infer the S\'{e}rsic exponent $\eta$ with an absolute error of $\leq$ 0.2 even for shallow SBPs. As for PSF-degraded data, \ifit\ can recover the input \SL\ model parameters with a satisfactorily accuracy almost over the entire considered parameter space as long as ${\rm FWHM(PSF)} \leq \mathrm{R}_{\rm eff}$. 
This study also includes examples of applications of \ifit\ to ETGs and local low-mass starburst galaxies. These tests confirm that 
\ifit\ shows little sensitivity on PSF corrections and SBP limiting surface brightness, and that subtraction of the 
best-fitting \SL\ model in two different bands generally yields a good match to the observed radial color profile.}
{It is pointed out that the publicly available \ifit\ offers an efficient tool for the non-supervised structural 
characterization of large galaxy samples, as those expected to become available with Euclid and LSST.}
\maketitle


\parskip = \baselineskip

\section{Introduction \label{intro}}
The S\'ersic fitting law \citep[\SL; also referred to as generalized de Vaucouleurs law,][]{Sersic63,Sersic68} is a widely used functional form in structural studies of galaxies, and intimately linked to our understanding on several key astrophysical subjects, such as the nature of bulges \citep[e.g.,][for a review]{Davies88} and the size growth of early-type galaxies (ETGs) since $z \sim 2$ \citep[e.g.,][]{Trujillo06_size_evolution,Buitrago17}. 
Following initial applications of it to surface brightness profiles (SBPs) of ETGs \citep{Caon93}, 
this functional form is now implemented in several state-of-the-art 1D and 2D surface photometry packages\footnote{Examples include GIM2D \citep{Simard98-GIM2D,Simard02-GIM2D}, {\sc Galfit} \citep[][and extensions of it, e.g., GALAPAGOS, Barden et al. 2012, MegaMorph, H\"au\ss ler et al. 2013, {\sc Galfit-Corsair} Bonfini 2014, and BUDDI, Johnston et al. 2017]{Peng2002-Galfit,Peng2010-Galfit}, BUDDA \citep{deSouza04-BUDDA}, GASP2D \citep{MA08}, {\sc Archangel} \citep{SchombertSmith12}, IMFIT \citep{Erwin15-IMFIT},  {\sc Profiler} \citep{Ciampur16} and {\sc ProFit} \citep{Rob17}.} and routinely applied to galaxy images as a whole \citep[e.g.,][]{Hoyos11,vdW12,Guo13-CANDELS,Lange15-GAMA} or in bulge-disk decomposition studies \citep[e.g.,][]{Andredakis95,CdJB96,Gadotti09}.

Its simple, three-parameter functional form, 
that defines a SBP at the photometric radius \rr\ (\arcsec) as  
\begin{equation}
\rm \mu(R^{\star}) = \mu_0 + \frac{2.5}{\mathrm{ln(10)}}\,(R^{\star}/\alpha)^{1/\eta} ,
\label{eq1}
\end{equation}
with $\mu_0$ (\sbb) and $\alpha$ (\arcsec) denoting the central surface brightness and an angular pseudo-scale length, and the shape parameter $\eta$ regulating the profile curvature offers a versatile means for approximating a wide variety of galaxy profiles, from those of nearly exponential type~I galactic disks ($\eta\approx 1$) to the centrally steeply increasing SBPs of massive ETGs, with the specific case of $\eta\approx4$ corresponding to the \citet{dV48} law. 
Another popular formulation of the \SL\ involves the mean surface brightness \meff\ at the effective radius \reff\ therein (cf. Eq.~\ref{eq2}); $\eta$ being coupled to these parameters through a normalization factor \citep[][for details]{Caon93,Trujillo01}.

Since the 
90ies, the \SL\ has attracted a great deal of interest, with several studies exploring its mathematical formulation and the couplings between its free parameters \citep[e.g.,][]{GrahamColless97,Trujillo01,GrahamDriver05,Gra05}, stellar-dynamical origin \citep[e.g.,][]{Marquez00,Cen2014}, intrinsic luminosity density distribution it corresponds to \citep[e.g.,][]{Cio91,P93,P96a,ChakrabartyJackson2009,BaesGentile11,BaesVanHesse2011}, and the connection between $\eta$
and light concentration indices \citep{GTC01}, and the Gini coefficient \citep{Lotz04,Cibinel13}.

Some examples echoing the importance placed on the \SL\ include the broad usage of $\eta$ as a discriminator 
between classical bulges and pseudo-bulges \citep[e.g.,][]{Gadotti09,FisherDrory08,FisherDrory11,Neumann17} and proxy to the 
bulge-to-total (B/T) ratio \citep[][]{Lang14}, bulge morphology \citep[e.g.,][]{Fabricius12,Lange15-GAMA}, central velocity dispersion \citep{Graham02} and super-massive black hole mass \citep[SMBH;][]{Graham01,Savorgnan16}.
The importance of the \SL\ in our understanding on galaxy taxonomy and scaling relations is also reflected in the debate 
on the classical giant vs. dwarf galaxy dichotomy \citep[established mainly from fitting King models to SBPs, cf. e.g.,][]{BinggeliCameron91,BinggeliCameron93} that vanishes on the $\mu_0$ vs. $\eta$ plane, suggesting that dwarf ellipticals (dEs) are essentially the lower-luminosity extension of bright ETGs \citep{GrahamGuzman03,Graham11a}. 
Likewise, the trend for increasing $\eta$ with increasing optical ETG luminosity -- already apparent from early photographic work by \citet{BinggeliCameron93} and the analysis by \citet{Caon93} and  -- has prompted the proposal of it being a useful extragalactic distance indicator \citep[][see, however, Binggeli \& Jerjen 1998]{YC94}. Moreover, the \SL\ was found to yield a good approximation to the stellar mass surface density (\sstar) profiles of galaxy bulges since $z\sim 2$ \citep{Lang14}, which underscores its versatility as a tool for the systematization of galaxy physical properties.

Despite the indisputable virtues of the \SL, some lines of evidence suggest caution in the way its shape parameter $\eta$ 
is derived and interpreted in the evolutionary context of galaxies. One limitation stems both from its mathematical nature and the specifics of its fitting to galaxy SBP data points via standard $\chi^2$-minimization \citep[e.g., through the Levenberg-Marquardt 
nonlinear fitting algorithm][]{Levenberg44,Marquardt63}. 
For instance, \citet{GTC01} demonstrate that inherent to the mathematical nature of the \SL\ is a correlation between $\eta$ and \reff\ (cf. Eq.~\ref{eq2}). This parameter coupling was also described from an empirical point of view in \citet{Noe03} as a degeneracy between $\eta$ and the pseudo-scale length $\alpha$ in Eq.~\ref{eq1}.
As the latter authors and \citet{Cairos03} remark, fitting Eq.~\ref{eq1} to a genuine (perfect) S\'ersic profile robustly recovers its three parameters, almost regardless of the radius (or $\mu$) interval considered. However, in the case of an imperfect S\'ersic profile (\iSP), that is a SBP that strictly follows the \SL\ over only a limited interval in \rr, the fit can substantially depend on the radius (or $\mu$) interval considered. 
This rather obvious fact has actually non-trivial implications regarding the uniqueness of the best-fitting \SL\ model  
and its applicability to the structural characterization of galaxies.

One may enumerate several examples of local galaxies with \iSP\ light distributions -- both mono-component systems with a nearly spatially uniform stellar mass-to-light ratio (\mlr; e.g., ETGs), as well as composites of young and old stellar populations of differing spatial extent that mimic a high-$\eta$ S\'ersic profile. A third class of pseudo-\SL\ profiles can result from the excitation of an extended nebular halo around a compact starburst or active galaxy \citep{PapOst12}. 
Whereas such extreme objects are scarce in the local universe, they are presumably ubiquitous at high redshift, that is, in the epoch when most galaxies have experienced the dominant phase of their build-up \citep[e.g.,][]{Stark09,SdB10,Steidel14}.

An example for the first type of \iSPs\ is given by massive luminous ($M_V \la -21.5$) ETGs with a ``depleted'' core, prompting the need for a modified \SL\ function \citep[e.g., the 6-parameter core-\SL\ function proposed in][see also Trujillo et al. 2004]{Graham03coreSL}. A precise quantification of these central deviations from the \SL\ is of considerable interest in the light of their documented connection with the ETG radio and X-ray power \citep{Bender1989} that could hold clues to the buildup history and nuclear energy sources of these systems. For instance, the ``light deficit'' relative to inward extrapolation of the \SL\ model, if due to core evacuation by a binary SMBH \citep{Begelman80,Ebisuzaki91,Merritt06}, could be used to constrain the SMBH mass and yield insights into the growth of SMBHs and their interaction with the galaxy host, as well as on the role of mergers on the buildup of ETGs \citep[e.g.,][and references therein]{Bonfini18}. 

A similar type of \iSPs\ is presented by a subset of dEs with exponential profiles with a flat core
\citep[type~V SBPs in the notation by][]{BinggeliCameron93}\footnote{Such centrally flattening exponential SBPs 
were also found in the underlying host of several BCDs \citep{P96a,Noe03} and hypothesized to arise from its adiabatic expansion in response to starburst-driven mass loss \citep{FT89,P96b}.}. 
While this kind of SBPs can, in first order, be approximated by a single \SL\ with a $\eta$ $<$ 1, 
the best-fitting set of $\eta$ and $\alpha$ (or \reff) depend on the $\mu$ interval considered: fits to shallow images being restricted to the core-dominated region typically yield a $\eta$ $\la$ 0.5, whereas deeper images, capturing the outer exponential intensity fall-off yield a $\eta$ $\sim$ 1 \citep[e.g.,][]{Noe03}.

A significant dependence of the fit on the $\mu$ range considered is also to be expected when a single \SL\ model is applied to an 
\iSP\ galaxy with an additional luminosity component, as for example a compact bulge or bar, circumnuclear star formation (SF) ring or the sharp central luminosity excess of nucleated dEs \citep{BinggeliCameron91,DePropris05} or Seyfert galaxies \citep[e.g.,][]{Xanthopoulos96,SMM10}. 
For instance, \citet{Balcells03}, point out from a combined analysis of SBPs from ground based and Hubble Space Telescope (HST) data that smearing of a central point source with the adjacent emission in a bulge of $\eta$ = 1.7 can lead to an artificial increase of $\eta$ to $\sim$ 4, mimicking a de Vaucouleurs profile. Likewise, \citet{MA08} and \citet{Breda-MSc} find that decomposition of a barred 
late-type galaxy (LTG) solely into a bulge and a disk (i.e., omitting the bar) can strongly bias the best-fitting \SL\ model for the bulge. More generally, errors arising from an incomplete or inadequate 1D or 2D parametric decomposition scheme can be systematic, therefore potentially of greater concern than formal fitting uncertainties.

In all these cases of \iSPs, prior knowledge of the photometric structure of a galaxy facilitates integration in an image  decomposition scheme of the parametric ingredients needed for an exact modeling of a galaxy, including the luminosity fraction owing to a bona fide \SL\ component (as an example, the parametrization of a nucleated type-V dE as due to the superposition of a core-\SL\ model plus a central Gaussian). This is a rather theoretical option, however, at least in what concerns compact (or higher-$z$) galaxies: structural details in the centers of these systems are rarely visually accessible due to limited spatial resolution and smearing
with the point spread function (PSF), and can only be obtained after a multi-stage photometric analysis (inspection of fitting residuals and color morphology, unsharp masking, wavelet decomposition or \citet{Lucy74} deconvolution). 
As a result, a suitable adaptation of SBP decomposition schemes is impractical in studies of individual galaxies and prohibitively complex in the case of automated structural studies of large galaxy samples, which by necessity mostly employ a single \SL\ to model galaxy images \citep[e.g.,][]{Simard98-GIM2D,MarleauSimard98,Griffith12,vdW12}.

Quite importantly, since possible deviations from the \SL\ in central parts of galaxies involve precisely those pixels with the highest signal-to-noise ratio (\snr) and therefore smallest photometric uncertainties $\sigma_{\mu}$, they could have a strong and non-easily predictable impact on any error-weighted image decomposition (for instance, bulge-disk decomposition into two \SL\ components, one of them with a fixed $\eta$ = 1 
that accounts for the disk). This problem is further aggravated by the fact that the central low-$\sigma_{\mu}$ data points are 
most strongly affected by PSF convolution effects.
As pointed out in \citet[][see also Ribeiro et al. 2016]{P96a}, the fact that in error-weighted profile decomposition the solution is driven by these innermost (lowest-$\sigma_{\mu}$) data points may lead to a systematic failure of the model to describe the lower-surface brightness (LSB) periphery of galaxies: in bulge-disk decomposition studies, preference to the innermost points can lead to an artificial ``compactification'' of the disk (overestimation of its central surface brightness together with underestimation of its exponential scale length), which in turn results in the underestimation of the luminosity of the bulge. 

The second category of \iSPs\ is due to the superposition of evolutionary and spatially distinct stellar populations, and rather frequent among local star-forming galaxies, such as blue compact dwarfs \citep[BCDs;][]{LooseThuan1986a,P96a,Cairos01,GdP03}. Except for very few cases (see below) starburst activity in these systems takes place in one or several knots embedded within a more extended old elliptical stellar host. BCDs showing a central confinement of SF \citep[$\sim$ 20\% of the local BCD population, classified as nuclear-elliptical according to the scheme of][]{LooseThuan1986a} typically exhibit high-$\eta$ (3-4) \iSP\ SBPs \citep[e.g., \object{Haro 2} and \object{Haro\,3},][]{LooseThuan1986b,P96a}, reaching in some cases a $\eta > 6$ \citep{BergvallOstlin2002}. Given the strong radial color gradients \citep[up to $\sim$ 2 $B$-$R$ mag/kpc out to 1--2 host galaxy exponential scale lengths $\alpha$; cf.][]{P96b,P02}, the best-fitting $\eta$ for these systems can substantially depend on the photometric passband and the radius interval considered in the fit. Without multi-band photometry or spectroscopic information, BCDs and their ``cohorts'' \citep[e.g., compact narrow emission-line galaxies or green peas (GPs); cf.][]{Koo94,Cardamone09,Iz11,Amorin12}, being omnipresent at higher-$z$'s, are therefore solely on the basis of their high $\eta$ hardly distinguishable from compact passive ETGs.

Finally, an even more extreme class of \iSPs\ -- exceptionally rare in the nearby universe yet probably omnipresent at high $z$'s -- are those presented by compact galaxies with a very high specific star formation rate (sSFR), such as the extremely metal-poor 
(12 + log(O/H) $\approx$ 7.2) BCDs \object{I Zw 18} \citep{SS70,Iz01} and \object{SBS 0335-052\,E} \citep[e.g.,][]{Iz90,P98,Herenz17}.
These systems experience a vigorous starburst episode that gives rise to a large ionized gas envelope reaching out to several kpc away from their stellar component. Since the nebular halo of starburst galaxies shows a nearly exponential H$\alpha$
profile \citep{P02,Knollmann05}, its superposition with the more compact stellar emission results in 
a two-slope exponential profile that is barely distinguishable from a high-$\eta$ S\'ersic profile, as discussed in \citet{PapOst12}. 
As shown by these authors, fitting a \SL\ to this kind of \iSP\ profiles can lead to an increasing S\'ersic exponent $\eta$ with decreasing limiting surface brightness \mulim, from $\eta$ $\simeq$ 1 when shallow imaging allows only for detection of the inner (stellar emission dominated) exponential part of the SBP to $1 \leq \eta \leq 5$ when deeper imaging data additionally enable detection of the surrounding shallower exponential nebular halo. Therefore, depending on the rest-frame \mulim\ of the imaging data in hand, such extreme starburst galaxies at high-$z$ could readily be misclassified as massive ETGs on the basis of their high-$\eta$ pseudo-S\'ersic profiles\footnote{Due to severe ionized gas contamination of these sources, and depending on their redshift and available photometric passbands, their observed colors could superficially support this erroneous conclusion \citep[cf. discussion and Fig. 15 in][]{PapOst12}.}.
This obviously applies to any system hosting a powerful central source of energy and momentum that is capable of exciting an extended nebular envelope, such as quasars. Moreover, the fact that escaping and resonantly scattered Lyman-$\alpha$ radiation, both in quasars \citep{Steidel11,Borisova16,Wisotzki16,VM18,Ari18,Wisotzki18,Can18} and starburst galaxies \citep{Hayes07}, leads to 
nearly exponential rest-frame UV halos suggests that \iSPs\ similar to those of \object{I Zw 18} could be ubiquitous 
among galaxies in the early universe.

Summarizing, whereas the S\'ersic law arguably gives a good first-order match to a wide range of observed galaxy SBPs and offers   
a convenient means for their quantitative characterization, it is actually only a subset of local galaxies (e.g., a part of ETGs) 
that display genuine (perfect) \SL\ profiles in all optical-NIR photometric bands and across their entire radial extent.
In fact, a large number of galaxies in the local universe, and the more so at high $z$'s, shows appreciable-to-strong deviations 
from the \SL. The modeling of such imperfect S\'ersic-like SBPs can strongly depend on the specifics of fitting, 
correction for PSF convolution effects and depth (\mulim) of the data in hand. These fitting uncertainties pose a significant 
obstacle to the automatized application of the \SL\ to large galaxy samples and can propagate into significant scatter
in fundamental galaxy relations.

In view of such considerations it appears worthwhile to explore a robust and computationally inexpensive \SL\ fitting concept that 
can be applied in an unsupervised manner \citep[e.g., without need for an initial guess to the \SL\ model parameters, like in \galfit, 
or a prior `training' on local galaxies, like in the case of convolutional neural networks, e.g.,][among others]{HC15} and 
having little sensitivity to $\mu_{\rm lim}$ and PSF convolution effects. Such a tool would be valuable to the automated structural characterization of the vast number of high-$z$ irregular \iSP\ galaxies expected to be detected with, for instance, the Euclid satellite and the Large Synoptic Survey Telescope (LSST). 

The fact that the galaxy morphology drastically changes toward high-$z$'s, with an increasing fraction of, for example, clumpy disks \citep{Elm09,Wis13,Shib15,Guo15-UV-clumps} and cometary (also referred to as tadpole) and ``chain'' galaxies \citep{Elm07,Straughn06} underscores the need for fitting the \SL\ without prior assumptions on morphology and structure (e.g., the B/T ratio or Gini coefficient) that are largely empirically established from analysis of lower-$z$ galaxy samples.

In this article, we present such a concept, which is realized in the publicly available code \ifit\footnote{Available at www.iastro.pt/research/tools.html, along with instructions for its usage, as well as observed and synthetic images for demonstration and test purposes.} and validated both on local Hubble-type galaxies \citep[][and this study]{BP18,Breda-PhD} and irregular star-forming galaxies, such as BCDs. A distinctive feature of \ifit\ is that the fit to a galaxy is not determined through standard error-weighted $\chi^2$ minimization with respect to individual SBP data points, but instead through the search for the best-fitting equivalent \SL\ model (hereafter eSP) that gives the best match both to the observed light growth curve and variation of the mean surface brightness \mmu\ with \rr. 
Basing the fitting procedure on the latter robust quantities facilitates quick convergence to a unique solution 
both for perfect and imperfect S\'ersic profiles without the need for an initial guess to fitting parameters. 
Quite importantly, a key advantage of \ifit\ over standard \SL\ fitting tools is that the solution is not driven 
by the innermost (\rr $\la$ FWHM) data points that enclose a typically minor fraction ($<$ 10\%) of the total luminosity 
and can strongly suffer from PSF degradation. \ifit, supplemented with the capability of retrieving \SL\ model parameters even from strongly PSF-degraded galaxy images offers a handy tool for the structural characterization of galaxies (or sub-components thereof, e.g., bulges) near and far.

In Sect. \ref{meth} we outline the concept of \ifit\ and in Sect.~\ref{howto} we provide a brief description of its invocation and output. A quantitative assessment of the ability of \ifit\ to retrieve the S\'ersic model parameters from synthetic SBPs (both unconvolved and convolved with a PSF model) spanning a relevant range in $\eta$ and $\mu_{\rm lim}$ is given in Sect.~\ref{iFIT_test}.
This section also provides examples of the application of \ifit\ to irregular starburst galaxies, exemplified by the BCDs \object{He 2-10} and \object{I Zw 18}. Finally, Sect.~\ref{conc} summarizes the main conclusions from this study, Appendix A gives a description of results obtained with \ifit\ for a sample of local ETGs and two distant ($z$ = 0.768 \& 1.091) galaxies and Appendix B provides an empirical rationale for the advantage of adopting alternative weights when performing weighted least squares to retrieve the correct parameters in the case of \iSP's.\\
At this stage \ifit\ is merely a standalone code that can be applied to any SBP provided in ASCII format. It is intended as a module of a versatile galaxy decomposition package under development that will offer a suite of functional forms, including a core-\SL\ function \citep{Graham03coreSL}, a centrally flattening modified exponential function \citep{P96a}, Ferrer profiles, the Nuker law \citep{Lau95}, and King profiles \citep{Els99}, among others, for approximating galaxy SBPs and their structural components.
For convenience, the \ifit\ distribution package is supplemented by a python tool that allows for the computation of SBPs with the algorithm by \citet{BenMoe87}.
\begin{figure}[h] 
\centering
\makebox[\textwidth][c]{\includegraphics[width=1\linewidth]{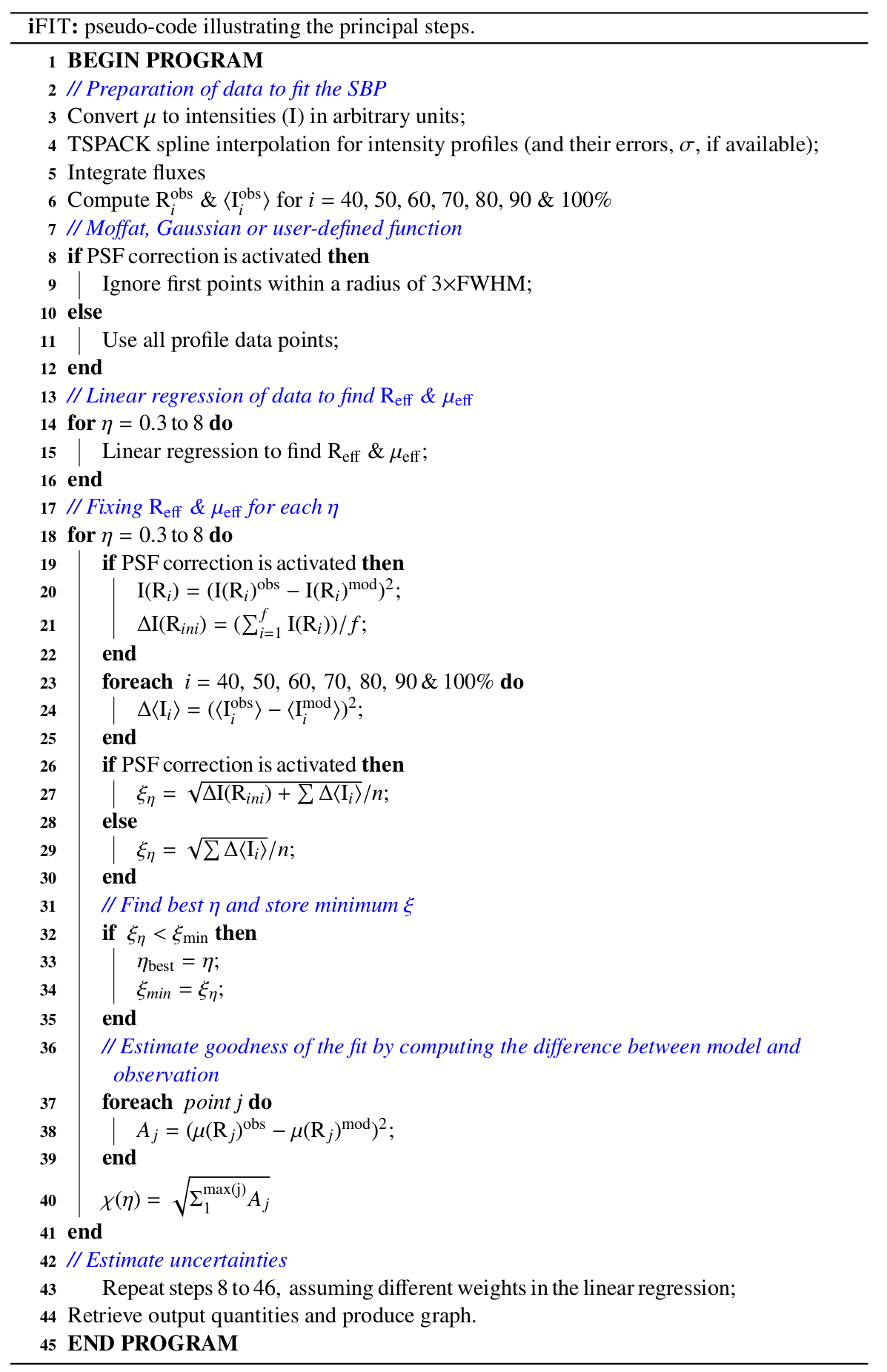}}
\caption{Pseudo-code illustrating principal steps of the convergence procedure in \ifit.}\label{ifit_pc} 
\end{figure}

\section{Methodology \label{meth}}
A widely used alternative formulation of the \SL\ is:

\begin{equation}
\rm \mu(R^{\star}) = \mu_{\rm eff} + \frac{2.5 b_{\eta}}{ln(10)} \left[ \left( \frac{R^{\star}}{R_{\rm eff}}\right)^{1/ \eta} - 1 \right],
\label{eq2}
\end{equation}
where, \reff\ is defined as the photometric radius that encloses 50\% of the total luminosity of a SBP, $\mu_{\rm eff}$ is the surface brightness at \reff\ and $b_{\eta}$ is given by $\Gamma(2 \eta) = 2 \gamma(2 \eta, b_{\eta})$, with $\Gamma$ and $\gamma$ standing for the incomplete and complete gamma functions, respectively \citep{Cio91}. The factor $b_{\eta}$ ensures that the radius \reff\ encloses 50\% of the total (out to \rr = $\infty$) luminosity of the SBP. At variance to the approximation $b_{\eta} = 1.9992 \eta - 0.3271$ proposed by \citet{Gra05} we adopted the asymptotic expansion given by Eq.~18 of \citet{CioBer99}.

Below, we summarize the main features of the adopted methodology: 

\begin{figure}[h]
\centering
\makebox[\textwidth][c]{\includegraphics[width=0.9\linewidth]{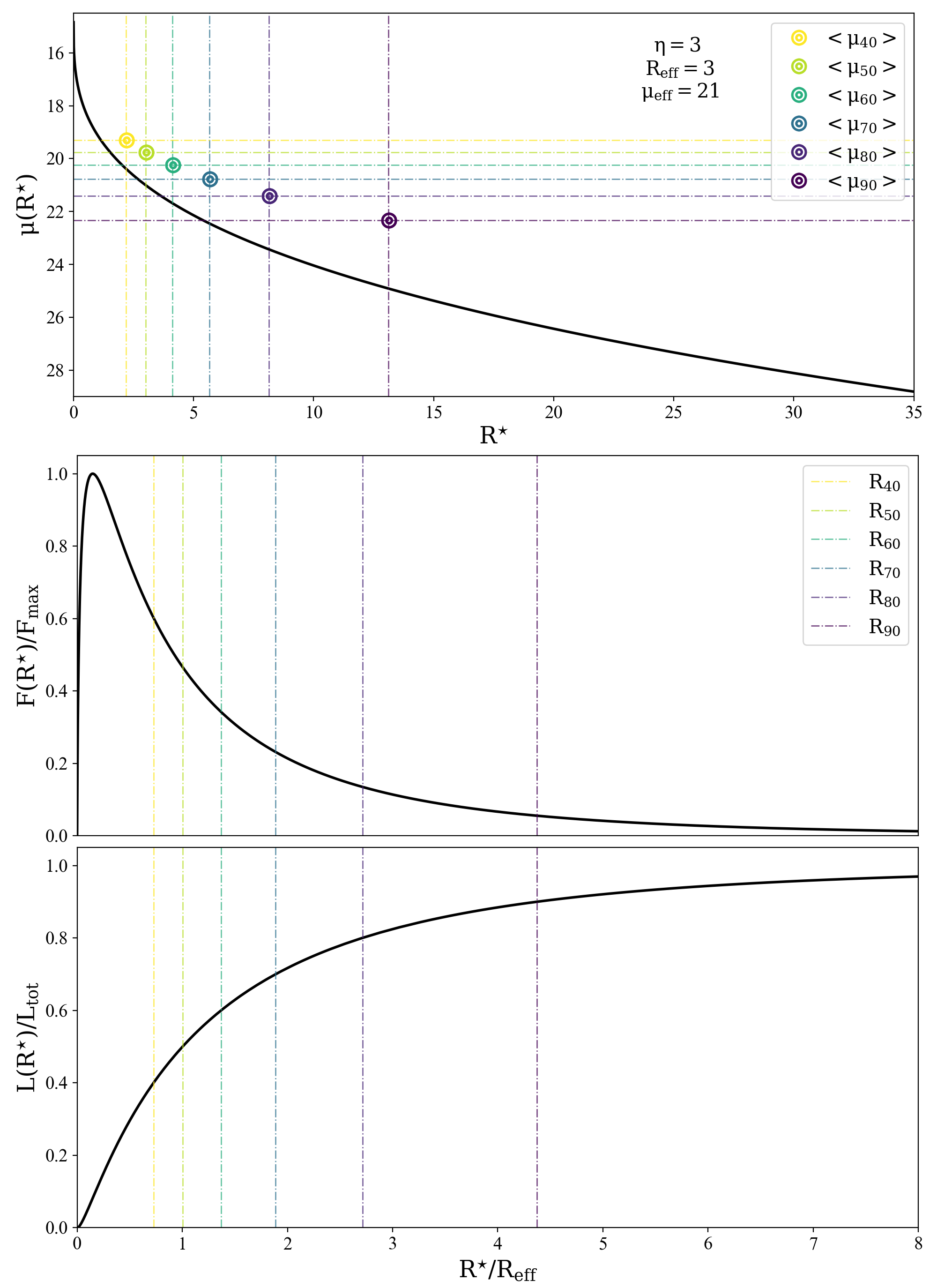}} 
\caption[Example of a \SL\ $\mu(\rm R^{\star})$ profile and respective flux distribution]{Example a S\'ersic profile 
$\mu(\rm R^{\star})$ with a $\eta$ = 3 and an effective radius \reff = 3\arcsec\ (solid line). \brem{Top panel:} Vertical dashed lines correspond to the radii enclosing 40, 50, 60, 70, 80 and 90\% of \lrmax, and horizontal lines depict the corresponding mean surface brightness \mmu\ therein. \brem{Middle panel:} Contribution to \lrmax\ as a function of \rr\ (i.e., integrand in Eq. 3) normalized to the maximum flux ($F(R^{\star})/F_{max}$. \brem{Bottom panel:} Light growth curve ${\rm L(R^{\star})/L_{tot}}$ of the profile. 
It can be seen that, although the innermost points (\rr/\reff $\leq$ 0.25) show the highest intensity, they are nearly irrelevant in terms of their contribution to the total luminosity.}
\label{SBP_flux} 
\end{figure}

\subsection{Preparatory steps \& Integration \label{sect2-1}}
\ifit\ starts by converting the input surface brightness profile ($\mu$, \sbb) into a radial intensity distribution $\rm I(R)$ (counts/$\sq\arcsec$), which is subsequently integrated out to the maximum observed radius 
$\mathrm{R}_{\rm max}$:

\begin{equation}
\rm L(R^{\star}_{\rm max}) = 2 \pi \int_{0}^{R_{\rm max}} I(R^{\star}) \cdot R^{\star} \cdot dR \\
\label{eq3}
\end{equation}

where $\rm I(R^{\star})$ is $10^{(\mu(R^{\star}) - C)/-2.5}$ and $C$ denotes a calibration constant in mag. Before integration, spline interpolation is performed by means of \tspack, a curve-fitting package based on exponential tension splines with automatic selection of tension factors \citep{tspack} that is publicly available for several programming languages. Due to the drastic steepening of high S\'ersic $\eta$ profiles at small \rr, traditional cubic spline interpolation techniques fail to accurately reproduce the \irr\ curve, which results in major errors in \lrmax. Whenever errors ($\sigma_{\mu}$) are provided along with an SBP, they are interpolated and taken into account in the fit.\\

\subsection{
Estimation of flux-enclosing parameters \label{sect2-2}} 
The latter permits estimation of a sequence of radii $\mathrm{R}_{40}$, $\mathrm{R}_{50}$, $\mathrm{R}_{60}$, $\mathrm{R}_{70}$, $\mathrm{R}_{80}$, $\mathrm{R}_{90}$ and $\mathrm{R}_{100}$ that enclose an increasing fraction of \lrmax\ (40, 50, 60, 70, 80, 90\% and 100\%, respectively) along with the corresponding mean intensity levels ($\langle \mathrm{I}_{40} \rangle$, $\langle \mathrm{I}_{50} \rangle$, $\langle \mathrm{I}_{60} \rangle$, $\langle \mathrm{I}_{70} \rangle$, $\langle \mathrm{I}_{80} \rangle$, $\langle \mathrm{I}_{90} \rangle$ and $\langle \mathrm{I}_{100} \rangle$) therein. The mean intensity $\langle \mathrm{I}_{i} \rangle$ corresponding to each \lrmax\ percentages ($i$) is computed as
\begin{equation}
\langle \mathrm{I}_{i} \rangle = \frac{i\cdot10^{-2}
 \cdot \mathrm{L}({\mathrm{R}^{\star}}_{\rm max})}{\pi \mathrm{R}_{i}^2}. \\
\label{eq4}
\end{equation}

As we will detail next, at variance to traditional \SL\ fitting techniques, \ifit\ determines the solution by evaluating the difference (model vs. observed SBP) for these $\langle \mathrm{I}_{i} \rangle$, this way preventing the few innermost points to dictate the fit. 
As apparent from the middle panel of Fig.~\ref{SBP_flux}, these central points with the highest intensity, which are most affected by PSF convolution effects, actually encompass a rather small fraction of \lrmax\ ($\la$ 10\%, depending on $\eta$).
Furthermore, the concept of \ifit\ secures a more robust solution against deviant SBP points due to, for instance, an incompletely removed overlapping Galactic star or a high-$z$ background source that pops up only in red passbands, since the latter have generally a small effect on \lrmax\ and the galaxy's light growth curve. 

\subsection{Determination of \reff\ \& $\mu_{\rm eff}$ \label{non-PSF}} 
One of the advantages of the presented algorithm is that in early stages of the convergence procedure it fixes two (\reff\ \& $\mu_{\rm eff}$) of the three free parameters of Eq.~\ref{eq2}, which in turn alleviates the degeneracy between \reff\ and $\eta$. This is done through a preliminary linear regression in $y = \mu(\mathrm{R}^{\star})$ vs. $x = \mathrm{R}^{1/\eta}$, where $\eta$ is allowed to vary from 0.3 to 8.0.
By employing least squares method for the functional form $\mu(\mathrm{R}^{\star}) = m \cdot \mathrm{R}^{1/\eta} + b$, one obtains 
for each adopted $\eta$ the effective radius \reff\ and the corresponding $\mu$ from the slope $m$ and the $y$-axis intercept $b$ as:

\begin{equation}
\mathrm{R}_{\rm eff} = \frac{b_{\eta}^{\eta}}{(m \cdot \mathrm{ln(10)}/2.5)^{\eta}} \quad {\rm and}\quad \mu_{\rm eff} = b + \frac{2.5 b_{\eta}}{\rm ln(10)}.
\label{eq4a}
\end{equation}

If a $\sigma_{\mu}$ array is provided by the user,  the previous step is repeated by performing weighted linear regression with the weight of each SBP point set to \brem{a}) $w(i) = 1/\sqrt{\sigma_{\mu}(i)}$, \brem{b}) $w(i) = 1/\sigma_{\mu}(i)$ and \brem{c}) $w(i) = 1/\sigma_{\mu}(i)^2$. Repeating the convergence procedure three times (obtaining three solutions) allows \ifit\ to estimate errors for the parameters of the \SL\ equation
\footnote{Appendix B contains a detailed explanation supporting the use of the modified weights.}.
For this reason we advice the user to supply $\sigma_{\mu}$ values in the 3$^{\rm rd}$ column of the ASCII file holding the SBP to be fit.

If the input SBP is convolved with a PSF, in order to ensure that the degraded resolution does not affect the determination of \reff\ and $\mu_{\rm eff}$, the linear regression is computed only for SBP data points within a radius \rr\ $>$ 3$\times$FWHM of the PSF. We note that in cases of strong PSF degradation (when FWHM/\reff\ $>$ 1-2) and a shallow input SBP that contains less than 90\% of the total luminosity (that is, a SBP truncated at the radius $R_{90} \simeq 4\times \mathrm{R}_{\rm eff}$ in the case of the example in Fig.~\ref{SBP_flux}), \ifit\ might not be able to obtain a sensible solution (see Fig.~\ref{fig3} for acquaintance of \ifit\ limitations). 
This should not be perceived, however, as a limitation that is exclusive to \ifit, since any surface photometry code presumably fails under such conditions.

\subsection{Estimating the  S\'{e}rsic $\eta$ for non-seeing convolved profiles \label{sect2-4a}} 

This module integrates a fundamental and novel aspect of the concept presented here: its goal is to estimate $\eta$ through the search for the best match between the seven observationally determined \mmu's ($i$ = 40, 50, 60, 70, 80, 90 and 100) with the theoretically expected values. Basing the solution on the 7 aforementioned robust quantities is an important advantage as compared to standard $\chi^2$ minimization with respect to SBP data points, and results in a unique \eSP\ to any SBP, namely, both to a perfect and imperfect \SL\ profile. In the first case, the deviation between model and SBP data points is obviously zero. Specifically, convergence is achieved through comparison of the theoretical $\mathrm{R}_{i}$ and \mmu\ for a grid of synthetic \SL\ profiles in the range 0.3 $\leq \eta \leq$ 8.0, adopting the respective \reff\ and $\mu_{\rm eff}$ for each $\eta$ obtained in the previous step.
\begin{equation}
\renewcommand*{\arraystretch}{1.5}
\begin{array}{l}
\langle \Delta\mathrm{I}_{i}\rangle = (\langle \mathrm{I}_{i}^{\rm obs} \rangle - \langle \mathrm{I}_{i}^{\rm mod} \rangle)^{2} \\
\xi = \sqrt{\sum \langle \Delta\mathrm{I}_{i}\rangle} / n
\end{array}
\renewcommand*{\arraystretch}{1}
\label{eq5}
\end{equation}
where $i$ stands for a percentage of \lrmax\ and $n$ is the number of terms inside the square root. The $\eta$ that yields the closest match between the theoretical and observed quantities (the absolute minimum in $\xi$) is selected as the best-fitting \eSP\ 
to a SBP.
This set of multiple constraints yields a robust and unique solution for $\eta$, eliminating the risk of being trapped in a local minimum, as may occur with standard nonlinear fitting algorithms whose best-fitting solution might depend on the initial guess and the selection (or $\sigma_{\mu}$) of individual data points. 

\subsection{Estimating the S\'{e}rsic model parameters for seeing convolved profiles \label{PSF}} 

\ifit\ allows to recover the \SL\ parameters from a SBP that is convolved by a Gaussian or Moffat \citep{Moffat69} PSF. The latter is given by
\begin{equation}
\renewcommand*{\arraystretch}{1.5}
\begin{array}{l}
\mathrm{F}_{\rm M}(\mathrm{R}^{\star}) = \Bigg[ 1 + \Bigg( \dfrac{\mathrm{R}^{\star}}{\alpha_{\rm M}}\Bigg)^{2} \Bigg]^{- \beta_{\rm M}}\\[2ex]
\alpha_{\rm M} = \dfrac{\rm FWHM}{(2 \sqrt{2^{1/\beta_{\rm M}}-1}})
\end{array}
\renewcommand*{\arraystretch}{1}
\end{equation}
Additionally, \ifit\ allows for correction for a user-supplied PSF in ASCII format. In the latter case, a Moffat model is fit to that PSF and the parameters $\alpha_{\rm M} \, \& \, \beta_{\rm M}$ are determined with \minpack\ \citep{minpack}, a publicly available set of subroutines that solves nonlinear least-squares problems using a modified version of the Levenberg-Marquardt algorithm. 
If, on the other hand, the user does not have an observationally determined ASCII profile for the PSF, \ifit\ offers the alternative of
rendering the FWHM (\arcsec) of the PSF for a Moffat or Gaussian function (in the first case $\beta_{\rm M}$ is assumed to be 4.765 \citep{Tru01} and $\infty$ for a Gaussian).

Once an analytic model for the PSF has been defined, \ifit\ constructs, similar to the treatment of non-convolved profiles, a set of synthetic SBPs by varying $\eta$ and subsequently convolving each one with the PSF. The procedure to find the correct $\eta$ is the same as the one described under Sect. \ref{sect2-4a} with the only difference being that the quantities $\langle \Delta\mathrm{I}_{i}\rangle$ 
now refer to PSF-convolved synthetic SBPs. Another difference lies in Eq.~\ref{eq5} which in this case includes an extra term that facilitates quick convergence: as shown in Fig.~\ref{SBP_flux}, although the innermost points account for a small fraction of the total luminosity, they dominate in terms of intensity. This implies that, especially for high-$\eta$ \SL\ profiles,  $\langle \Delta\mathrm{I}_{40}\rangle$ and $\langle \Delta\mathrm{I}_{50}\rangle$ are similar for different $\eta$'s.
The introduction of an additional single term that compares the observed surface brightness at the innermost SBP point
with those of the model at the same radius eases convergence to the correct solution ($\Delta \mathrm{I}(\mathrm{R}_{ini})$ is defined as the sum of the differences between observations ($\mathrm{I}(\mathrm{R}_{i})^{\rm obs}$) and model ($\mathrm{I}(\mathrm{R}_{i})^{\rm mod}$) for the set of points that lay within the first observed point ($i = 1$) and point $i=j$ -- the first observational point where the derivative increases, that is, $f'(i)|_{i=j} > f'(i)|_{i=j-1}$).

\begin{equation}
\renewcommand*{\arraystretch}{1.5}
\begin{array}{l}
\delta \mathrm{I}(\mathrm{R}_{i}) = (\mathrm{I}(\mathrm{R}_{i})^{\rm obs} - \mathrm{I}(\mathrm{R}_{i})^{\rm mod})^{2}\\

\Delta \mathrm{I}(\mathrm{R}_{ini}) = (\sum\limits_{i=1}^{j}\delta \mathrm{I}(\mathrm{R}_{i}))/j\\

\xi = \sqrt{\Delta \mathrm{I}(\mathrm{R}_{ini}) + \sum \langle \Delta\mathrm{I}_{i} \rangle}/n
\end{array}
\renewcommand*{\arraystretch}{1}
\label{eq6}
\end{equation}

Finally, after convergence to a solution, \ifit\ computes the difference between model and observation providing an estimate for the goodness of the fit.

\section{Application of iFIT \label{howto}}
\ifit\ requires as minimum input the SBP of the galaxy in a two-column ASCII file holding the photometric radius \rr\ (\arcsec) and corresponding surface brightness $\mu$(\rr) (\sbb), with the option of including photometric uncertainties $\sigma_{\mu}(i)$ in a third column. If the SBP is to be modeled taking into account PSF convolution effects, the user needs to additionally provide the PSF in the same two-column format as the input SBP, or indicate the FWHM (\arcsec) of a Moffat or Gaussian approximation to the PSF. 
In its current public release, \ifit\ is provided along with an auxiliary routine that permits computation of SBPs through a python implementation of the 2D photometry code by \citet{BenMoe87}.

\ifit\ exports the results from the fit into three files, one of those holding a graphical representation of the input SBP and 
the \SL\ model of it in postscript (\textit{.ps}) format (see Appendix~\ref{appA} for a complete explanation). 
Of the two additional ASCII files, the first one (\textit{name\_of\_input\_file.SProf.dat}) stores the obtained \SL\ model 
in a two-column format for non-PSF-convolved SBPs and with four columns in the case of PSF-convolved data where the last two columns list the convolved \SL\ model. The second ASCII output file, \textit{name\_of\_input\_file.Results.dat}, contains the three parameters of the \SL\ best-fitting model ($\eta$, R$_{\rm eff}$ and $\mu_{\rm eff}$), apparent magnitude, absolute magnitude (whenever the distance to the galaxy is provided), central surface brightness $\mu(\mathrm{R}^{\star}=0)$ of the model as well as an estimate on the quality of the fit (0, 1 and 2 for, respectively, a adequate, moderate and inadequate fit).

A cookbook provided along with the distribution of \ifit\ (v.1) contains examples of the invocation of the code 
for different choices on the PSF (cf. Fig.~\ref{ifit-workflow}).
\begin{figure}[h]
\includegraphics[width=1.0\linewidth]{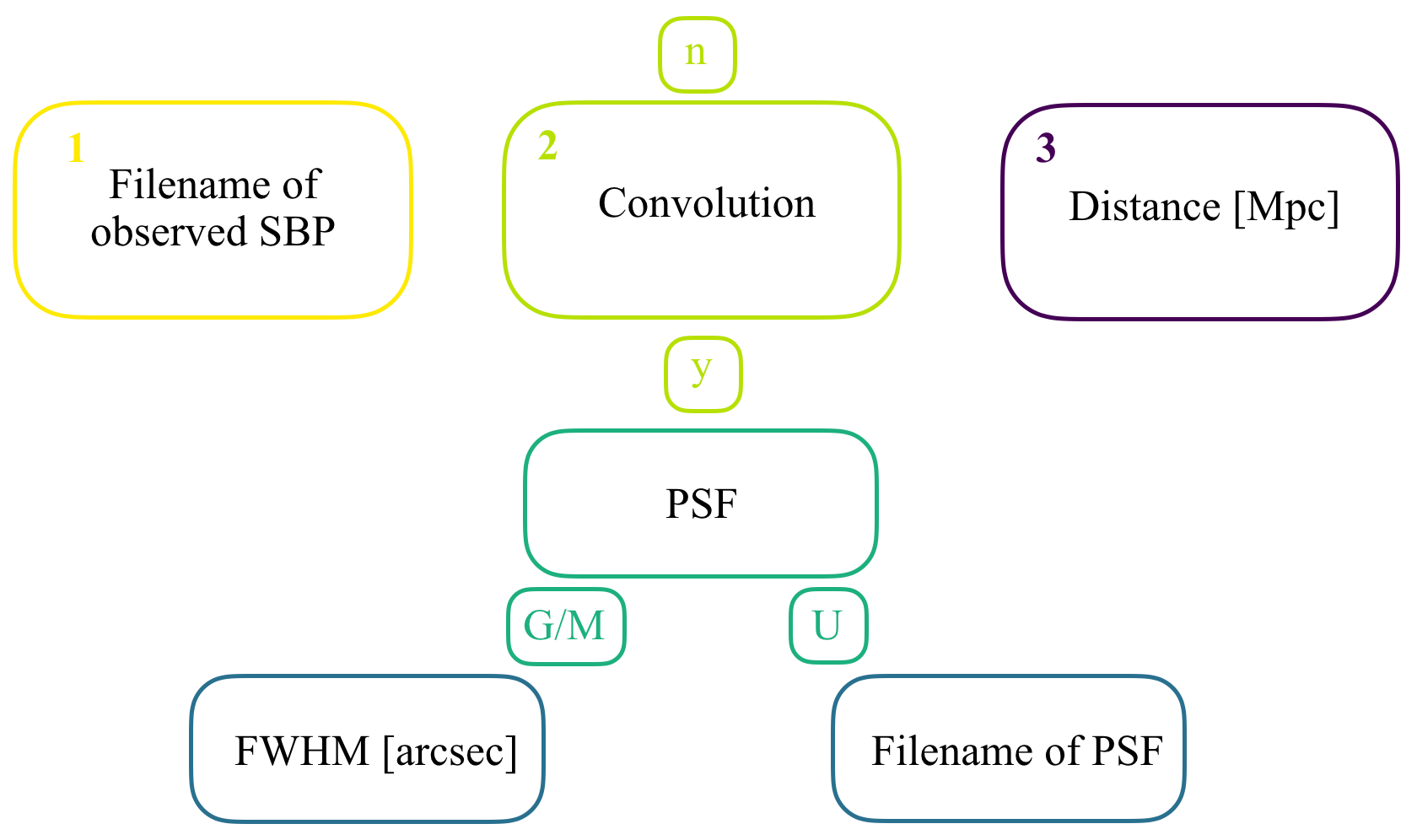} 
\caption{Scheme of input and output quantities, and available provisions for the PSF model to be taken into account, where ``y'' stands for ``yes'', ``n'' for ``no'', ``G/M'' for a Gaussian or Moffat PSF model, and ``U'' for a non-PSF-convolved SBP.
}\label{fig4a} 
\label{ifit-workflow}
\end{figure} 

Some examples are:

\begin{minipage}{8.6cm}
$\bullet$ fitting a SBP that is convolved with a Moffat PSF with a FWHM of 0.5 arcsec:
\begin{verbatim}
$ iFIT name_of_SBP_file.dat y M 0.5 0
\end{verbatim}
\end{minipage}%

\begin{minipage}{8.6cm}
$\bullet$ fitting a SBP that is convolved with a PSF provided by the user in ASCII format:
\begin{verbatim}
$ iFIT name_of_SBP_file.dat y U 
name_of_PSF_file.dat 0
\end{verbatim}
\end{minipage}%

\begin{minipage}{9cm}
$\bullet$ fitting the SBP of a galaxy at a distance of 10 Mpc without taking PSF convolution effects into account:
\begin{verbatim}
$ iFIT name_of_SBP_file.dat n 10
\end{verbatim}
\end{minipage}%
\section{Assessment of the accuracy of {\sc iFit} \label{iFIT_test}}
In order to assess the ability of \ifit\ to infer the S\'ersic model parameters we created an extended grid of synthetic profiles covering the relevant parameter range 0.3 $\leq$ $\eta$ $\leq$ 4.2 and 1 $\leq$ \reff\ $\leq$ 20 in steps of 0.1 and 1\arcsec, respectively. A second grid of synthetic profiles was created by convolving these SBPs with Moffat PSF models with a FWHM between 0\farcs5 and 2\arcsec\ in steps of 0\farcs5. Furthermore, these mock SBPs were truncated at the radius enclosing 90, 95 and 99\% of the total luminosity in order to simulate different limiting \murlim. The results are discussed in the following sections. 

\subsection{Testing iFIT on unconvolved synthetic profiles}   

\begin{figure*}[ht!]
\includegraphics[width=0.8\linewidth]{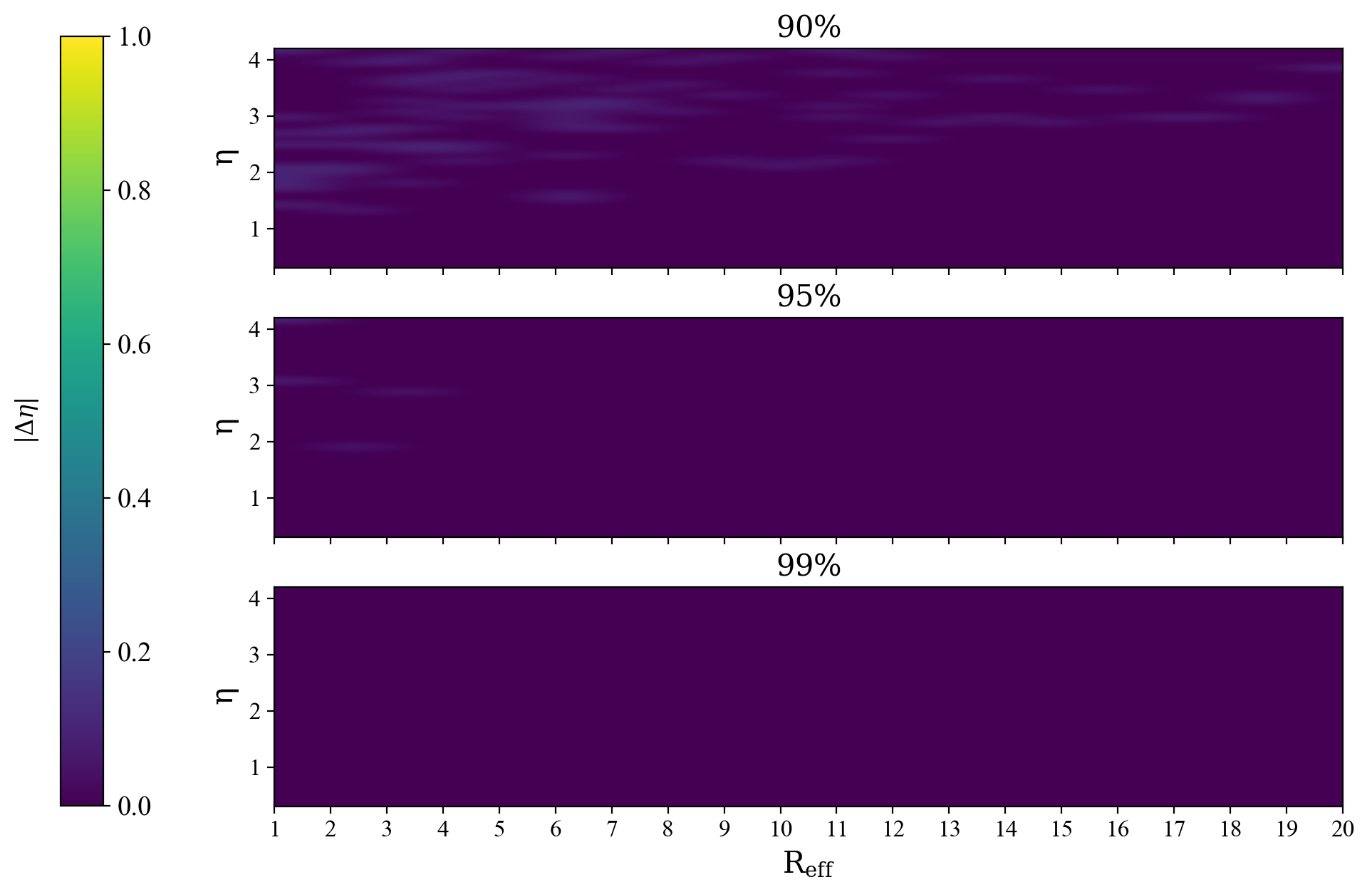}
\caption[\ifit\ $\eta$ error for unconvolved synthetic SBPs]{Absolute deviation $\Delta\eta$ between estimated and input $\eta$ as a function of $\eta$ and the effective radius \reff\ (\arcsec) for the case of unconvolved S\'ersic profiles encompassing 90, 95 and 99\% of the total luminosity (upper, middle and lower panel, respectively). The vertical color bar corresponds to $\Delta \eta = |\eta_{in} - \eta_{out}|$ between 0 and 1.}
\label{fig2} 
\end{figure*} 

Here we examine the absolute deviation $\Delta\eta$ between input and output $\eta$ for the studied range in $\eta$ and \reff. We consider three cases of non-PSF-convolved S\'ersic SBPs enclosing 90\%, 95\% and 99\% of \lrmax. It should be noted that for the case illustrated in Fig.~\ref{SBP_flux} truncation of the profile at $\mathrm{R}_{90}$ implies a limiting surface brightness \mulim $\la$ 25 \sbb\ that corresponds to a rather shallow exposure.
As apparent from Fig.~\ref{fig2}, for perfect S\'ersic profiles containing 99\% of the total SBP luminosity, the code recovers $\eta$ over the entire considered parameter range with no error, whereas fits to shallow (90\% \& 95\% of \lrmax) SBPs yield an error of typically less than 0.1 in $\eta$. 
Regarding the other two \SL\ parameters, \ifit\ recovers \reff\ \& ${\mu_{\rm eff}}$ with a maximum absolute error of 0\farcs4, 0\farcs01 and $\sim$0\arcsec\ \& 0.06 mag, 0.01 mag and $\sim$0 mag for profiles enclosing 90\%, 95\% and 99\% of \lrmax, respectively.
Given that the correct recovery of $\eta$ is tightly connected to that of the other parameters, with other words, failure to determine one of the three \SL\ parameters normally implies global failure of the fit, the topology of $\rm \Delta R_{eff}$ and $\rm \Delta \mu_{\rm eff}$ is similar to that of $\Delta\eta$.

\subsection{Testing iFIT on PSF-convolved synthetic profiles}   

The situation is obviously more demanding in the case of seeing-degraded S\'ersic SBPs, where the ability of any code to recover the input $\eta$ depends on the degree of PSF smearing.\\

\begin{figure*}[h!]
\includegraphics[width=0.8\linewidth]{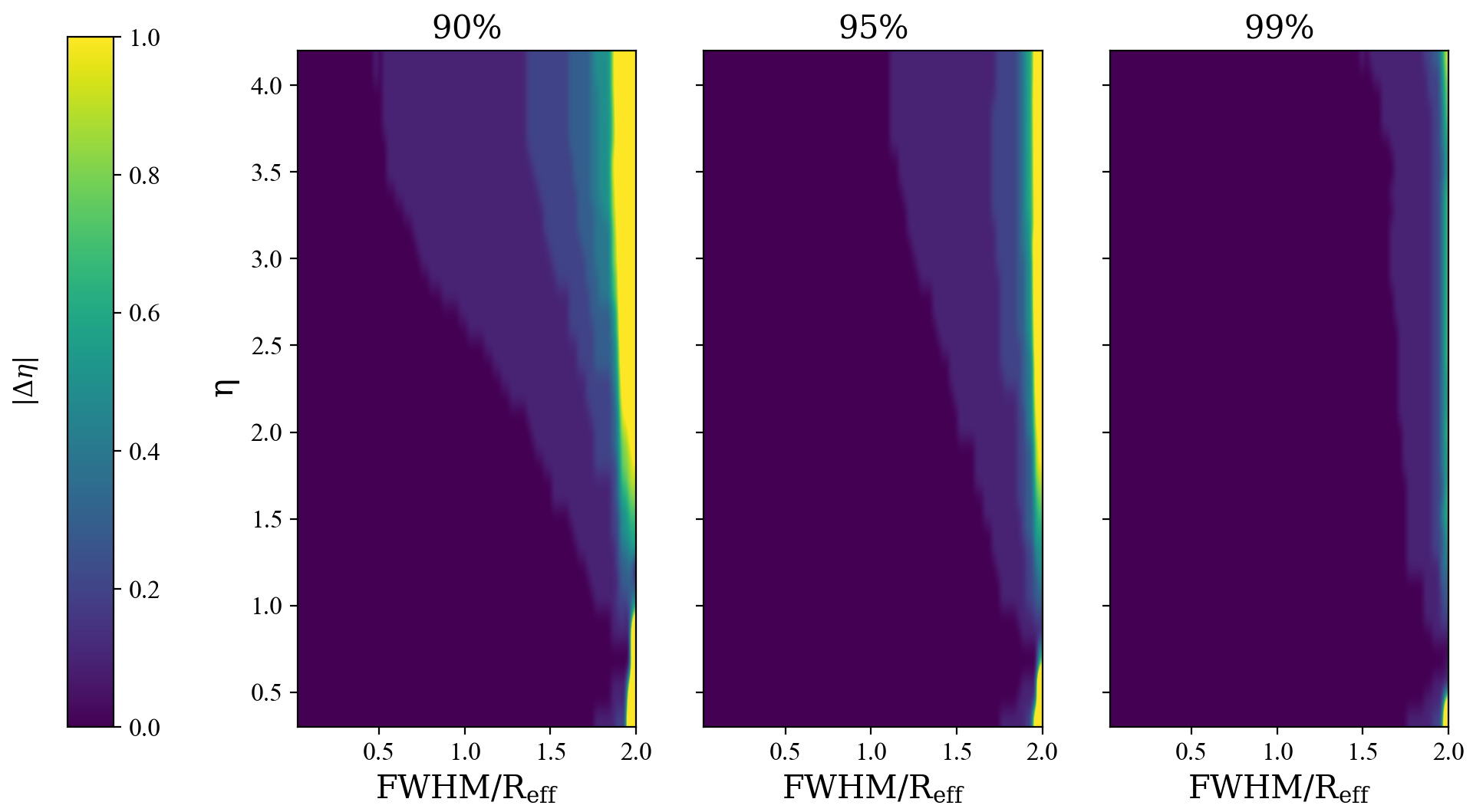}
\caption[\ifit\ $\eta$ error by testing using convolved synthetic data]{Absolute deviation $\Delta \eta$ between estimated and input $\eta$ as a function of $\eta$ and FWHM/\reff\ for the case of S\'ersic SBPs enclosing 90\% (left panel), 95\% (middle panel) and 99\% (right panel) of the total luminosity after convolution with a Moffat PSF with 0.5 $\leq$ FWHM/\reff\ $\leq$ 2, whereby 1 $\leq$ \reff\,(\arcsec) $\leq$ 20. 
}\label{fig3} 
\end{figure*} 
Figure~\ref{fig3} shows the results for synthetic SBPs covering the range 0.3 $\leq$ $\eta$ $\leq$ 4.2, 1 $\leq$ $\mathrm{R}_{\rm eff} (\arcsec)$ $\leq$ 20 and 0.025 $\leq$ FWHM/$\mathrm{R}_{\rm eff}$ $\leq$ 2. 

As apparent from Fig.~\ref{fig3}, \ifit\ faithfully recovers $\eta$ across the entire range of input values as long as the FWHM of the PSF does not appreciably exceed the profile effective radius. 
On the other hand, even for SBPs that contain 99\% of the total luminosity (right panel) there is a region in the parameter space where \ifit\ fails to recover the correct $\eta$ (and consequently also \reff\ and $\rm \mu_{\rm eff}$), especially but not exclusively for very low-$\eta$ profiles. 
This happens when FWHM/\reff\ $>$ 1, suggesting that in order for \ifit\ to converge into an accurate $\eta$, the PSF must be lower than \reff. This is actually a generous limit given that, for a FWHM/\reff\ $>$ 1, a structural analysis becomes increasingly problematic. 
For instance, in the case of a disk-dominated LTG this situation would imply that the FWHM of the PSF is equivalent to 1.7 exponential disk scale lengths.
In the even more extreme case of both strongly PSF-degraded and shallow SBPs, specifically, when $\rm R_{\rm max}$ is on the order of 3$\times$FWHM, \ifit\ is unsurprisingly unable to correctly estimate \reff\ and $\rm \mu_{\rm eff}$, therefore also $\eta$ (see Sect.~\ref{non-PSF}, item \textbf{3.}). 
Such a situation applies, however, to high-$z$ compact galaxies observed under natural seeing (FWHM $\sim$ 1.3\arcsec\ in the case of the Sloan Digital Sky Survey-SDSS) and for which the rest-frame \murlim\ is probably restricted due to cosmological dimming. If, however, higher-angular resolution data with, for example, the HST (maximum FWHM $\sim$ 0\farcs18) are available for such galaxies, \ifit\ permits an accurate ($\Delta\eta \, \leq$ 0.1) determination of $\eta$. 
As for $\rm | \Delta R_{eff} |$, in the range 0.025 $<$ FWHM/\reff\ $<$ 1 the maximal errors were found to be 0\farcs76, 0\farcs37 and 0\farcs34 for profiles enclosing 90\%, 95\% and 99\% of \lrmax, respectively. Regarding $\rm | \Delta \mu_{eff} |$, in the same range of FWHM/\reff, maximal deviations between input and retrieved values are 0.78 mag, 0.43 mag and 0.22 mag. 
As in the previous situation, the variation of the quantities $\rm | \Delta R_{eff} |$ and $\rm | \Delta \mu_{eff} |$ across FWHM/\reff\ is similar to that of $\rm | \Delta \eta |$. On the grounds of this error analysis, and from Fig.~\ref{fig3}, it is advisable to not use \ifit\ when the PSF (FWHM) is larger than the \reff\ of the observed profile, specially when the imaging data are shallow.

\subsection{Testing iFIT on early-type galaxies}\label{ifit-etgs}

\ifit\ was tested on $g$ and $r$ band images for 121 local ETGs from the SDSS (see Appendix~\ref{appA} for details). The images were sky-subtracted, rotated to the astronomical orientation, corrected for Galactic extinction following the prescriptions by \citet{SchFin11} and, whenever necessary, corrected for foreground Galactic stars by substituting them by the mean intensity of the adjacent stellar continuum. Finally, the frames were trimmed and SBPs were computed from 2D models obtained with the function FIT/ELL3 \citep[based on an algorithm by][]{BenMoe87} within the package SURFPHOT of ESO-MIDAS\footnote{We note that this code approximates a galaxy image as due to the superposition of thin elliptical annuli with a free center, position angle and ellipticity.

}.
Photometric uncertainties for SBP data points were computed following \citet[][hereafter P96a]{P96a}. Modeling with \ifit\ was performed by taking into account the PSF, as determined from typically three non-saturated Galactic stars in each frame. 

It was found that, although some PSF models yield a visually better fit at small radii, they generally play a minor role on the determination of the \SL\ parameters (see Fig.~\ref{fig1A} in the Appendix). 
This demonstrates the robustness of the algorithm. 
As an additional check, fits were repeated for 24 randomly selected ETGs from SDSS in the $r$- and $g$-band that were intentionally inaccurately sky subtracted, such as to leave residuals at the level up to $\sim$3 times the sky noise.
This test has shown that the result does not greatly depend on the quality of the background subtraction, given an average difference of 0.1 $\pm$ 0.11 in $\eta$ and 1\farcs6 $\pm$ 1\farcs62 in \reff\ as compared to the previously determined values.

Another key test has addressed whether subtraction of the best-fitting \SL\ model of SBPs in two different passbands yields a radial color profile that well matches the observed one. This was found to be indeed the case for the majority of the sample, as illustrated in Fig.~\ref{color_profs} for three randomly selected ETGs.
More specifically, a comparison of the derived $\eta$ for the ETG sample in the two filters yields an average difference 
|$\eta_{r}$ - $\eta_{g}$|/$\eta_{r}$ of 0.08. This value, and its small dispersion ($\sigma$ = 0.07) attests a smooth variation of $\eta$ across wavelength and is consistent with the similarity of the SBPs of these systems in both bands, as apparent from the very weak $g$--$r$ radial color gradients of the galaxies (Fig.~\ref{color_profs}) for \rr\ $\ga$ 3\arcsec. 
This is also due to the concept of \ifit, specifically, the fact that the \SL\ solution is constrained on the basis of robust 
\mmu's rather than by directly fitting the SBP data points. 
This an advantage over standard concepts: for instance, a study with \galfit\ of 17 massive galaxies from \citet{dR18} 
yields an average difference $|\eta_{H} - \eta_{I}|/\eta_{H}$ of $\sim$ 0.38 between solution in the $I$ and $H$ band, 
whereas for the same sample \ifit\ returns 0.22. 
The fact that standard fitting tools may give in some cases discordant $\eta$'s in different bands (which might be partly due to imperfections in the PSF models used) is a known issue that has been dealt with by different teams. For instance, the MegaMorph extension of \galfit\ \citep{H13-MegaMorph} integrates a prescription that forces $\eta$ to vary smoothly across wavelength within a narrow, empirically estimated interval. 
Such a mechanism is unnecessary in the case of \ifit, which in all studied cases yields a good match between predicted and observed radial color profiles.

\begin{figure}[h]
\includegraphics[width=1.0\linewidth]{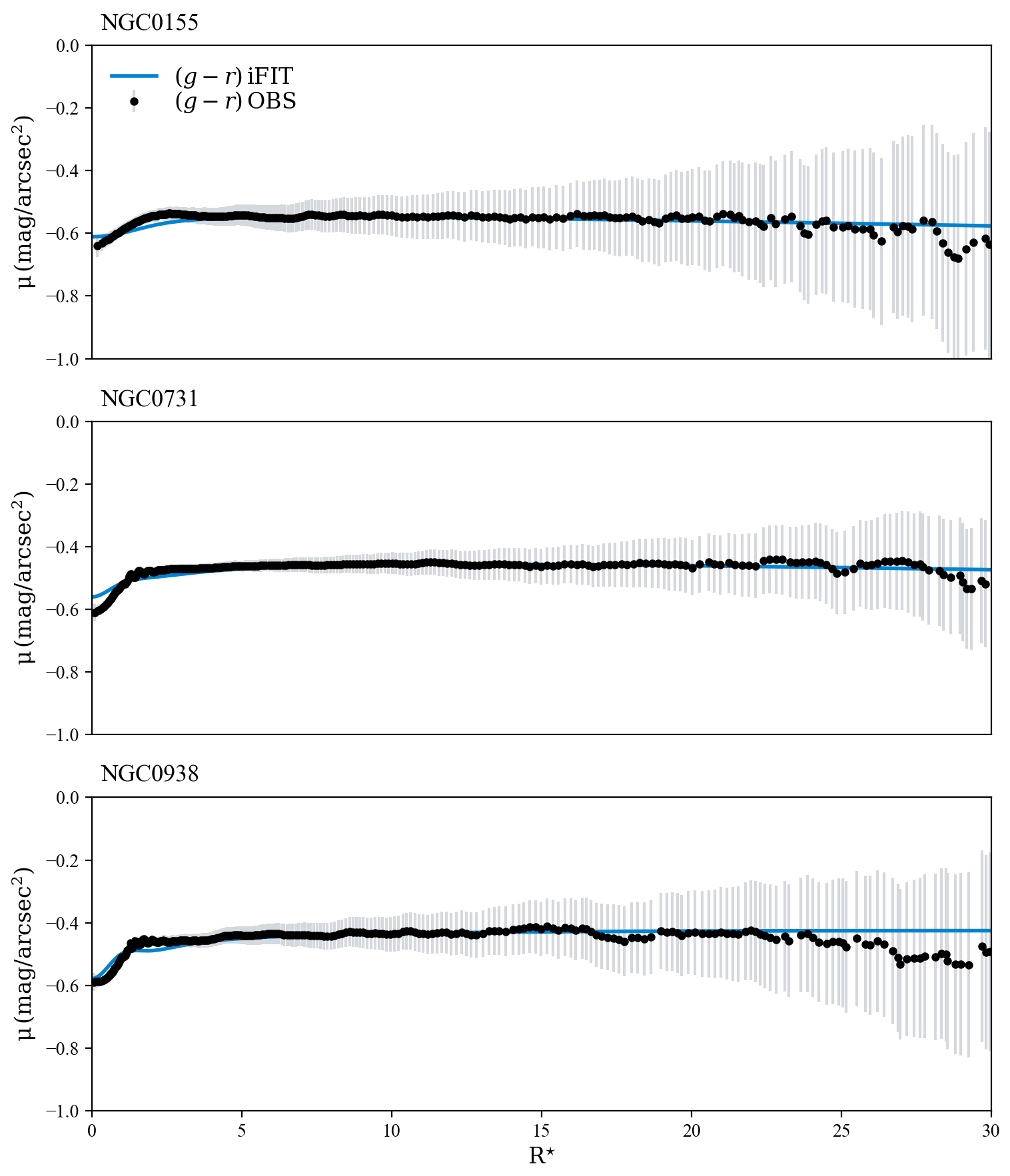} 
\caption{Observed $g$-$r$ radial color profile (filled circles) vs. difference between the best-fitting S\'ersic model with \ifit\ for $g$- and $r$-band images (blue curve) for three randomly selected ETGs from SDSS.}\label{color_profs} 
\end{figure}

\subsection{Application of iFIT to irregular starburst galaxies}
\ifit\ may have a wide range of applications on irregular galaxies, such as local BCDs/GPs and the majority of unevolved galaxies in the early universe. 
The structural analysis of such systems with standard tools poses a significant challenge given their clumpiness and asymmetry, the absence of a clear geometrical center and an axis-symmetric morphological pattern that could be easily integrated in a parametric decomposition scheme. 
The fact that the intensity maximum (i.e., the usually visually chosen ``galaxy center'') for such rapidly assembling systems may spatially differ across wavelength, due to SF propagation and strong detached nebular emission \citep[cf. e.g.,][for such cases among BCDs]{P96a,P98,P02}, as well as the redshift-dependent \mulim, due to cosmological dimming, further complicate the situation. A robust determination of the best-fitting 
\eSP\ model for such galaxies would be desirable in the context of an automated structural characterization of large galaxy samples with data from, for instance, Euclid and LSST. 
\begin{figure*}[h]
\includegraphics[width=1.0\linewidth]{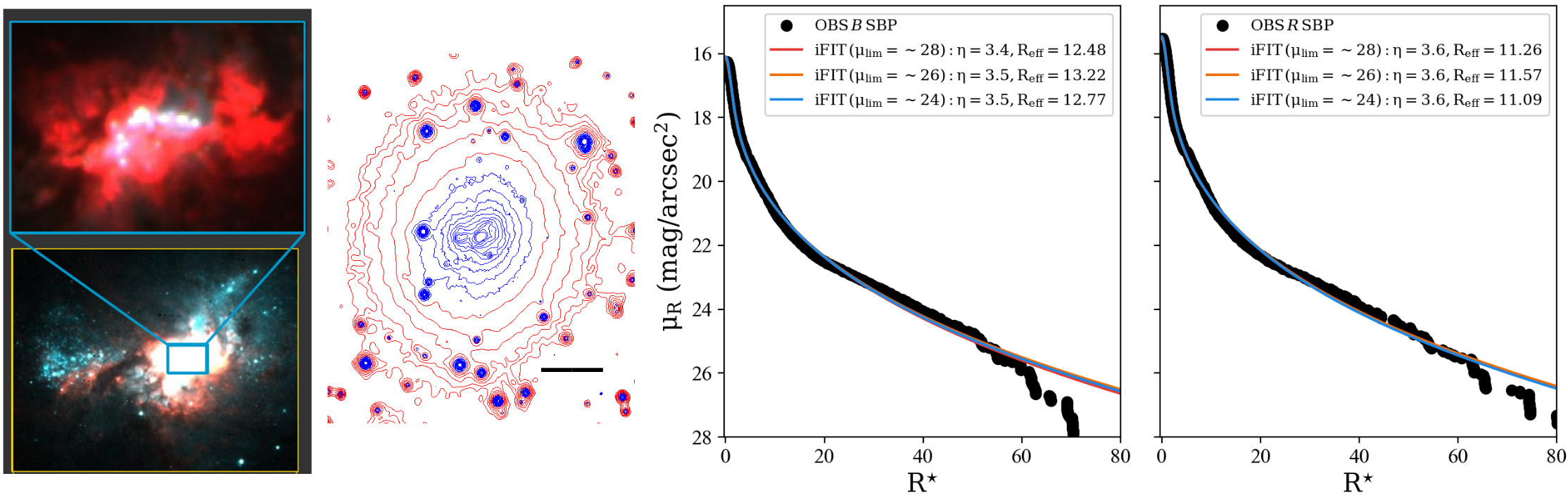}
\caption[]{\brem{Left panel:} True-color image composite of archival broad- and narrow-band HST images for the central region of the BCD \object{Henize 2-10}, illustrating the morphological complexity of stellar emission (green and blue) within its starburst component, as well as the presence of strong H$\alpha$ emission (red) excited by the chain of super-star clusters (zoom-in in the upper panel) discovered by \citet{CV94}. \brem{Middle panel:} Contour map of \object{He 2-10}, computed from ground-based data. The morphology of the starburst component and of the underlying older host galaxy is delineated by contours in blue and red, respectively. 
The horizontal bar corresponds to 20\arcsec.
\brem{Right panel:} $B$- and $R$-band SBPs of \object{He 2-10}, computed with the 2D surface photometry code by \citet{BenMoe87}. The overlaid curves show \SL\ fits with \ifit\ down to a surface brightness \mulim\ between 24 and 28 \sbb. It can be seen that in all cases the S\'ersic model parameters inferred for both bands are mutually consistent and show little dependence on \mulim.}\label{He210}
\end{figure*} 
Since, from the morphological point of view, local BCDs can be considered the best local analogs of unevolved low-mass galaxies at high-$z$, it is worth testing applications of \ifit\ to such systems.
As an illustrative example, we show in Fig.~\ref{He210} the \SL\ model obtained with \ifit\ for 
the BCD \object{Henize 2-10} by fitting $B$ and $R$ images taken with the Danish 1.54m telescope 
at La Silla \citep[see][for details]{PF98}. It can be seen that, \ifit\ yields for both filters similar S\'ersic model parameters, 
despite a difference by 4~mag in the adopted \mulim.
The stability of the best-fitting \SL\ solution with \ifit\ is to be contrasted with studies of BCDs with \galfit, which, as pointed out by \citet{Amorin07}, show a strong dependence on the surface brightness interval considered.

Another case is represented by \object{I Zw 18}, an extremely metal-poor BCD that is immersed within an extended nebular halo contributing $\ga$ 1/3 of its total $R$-band luminosity \citep{P02}. As shown in \citet{PapOst12}, the $R$-band profile of this galaxy (SBP1) reflects the superposition of two nearly exponential luminosity components differing in their scale length $\alpha$ and central surface brightness $\mu_0$, namely the steep inner component that is primarily due to stellar emission and dominates down to $\sim$ 24.5~\sbb, and the surrounding shallower nebular envelope. The best-fitting $\eta$ for the combined SBP depends on the limiting surface brightness, being $\sim$ 1.2 for a \mulim\ $\leq$ 24.5 \sbb\ and reaching an $\eta \sim$ 2 for a \mulim\ $\simeq$ 29 \sbb\ (dashed gray curve in Fig.~\ref{IZw18}). 
Would the nebular envelope present a shallower profile under preservation of its total luminosity (for instance, a by a factor 2 larger exponential scale length $\alpha$ and by 1.5~mag dimmer central surface brightness $\mu_0$), then the best-fitting $\eta$ for the combined stellar+nebular profile (SBP2) 
would increase from 2.5 for \mulim\ = 26 \sbb\ to up to $\sim$ 5 for \mulim\ $>$ 28 \sbb\ (dashed-dotted gray curve in Fig.~\ref{IZw18}). 
In the latter case, a marginally resolved distant morphological analog of \object{I Zw 18} (e.g., a high-sSFR compact galaxy or 
quasar surrounded by an extended Ly$\alpha$ envelope) could, purely on the basis of its high $\eta$, be misclassified as an evolved ETG.

Repeated modeling of SBP1\&2 with \ifit\ (Fig.~\ref{IZw18}) reveals a far weaker dependence of the 
\SL\ solution on \mulim: even for a \mulim\ $\simeq$ 29 \sbb, the inferred $\eta$ reaches maximal 
values of $\sim$ 1.7 and $\sim$ 2.5 for SBP1 and SBP2, respectively. These lower $\eta$'s 
better reflect the fact that $\sim$ 2/3 of the optical luminosity of \object{I Zw 18} originates 
from the steeper stellar emission-dominated inner exponential component.
Therefore, the concept of luminosity- and \mmu-weighted inference of S\'ersic model parameters in \ifit\ gives credit 
to those parts of a SBP that contribute most to the total emission while at the same time providing a good match to a SBP as a whole.
On the other hand, this fact alone does not mean that fitting an \eSP\
to such composite stellar + nebular profiles offers a meaningful approach to the structural characterization 
of compact, high-sSFR galaxies like \object{I Zw 18}. 
Supplementary or alternative approaches might be worth considering, as for example an analysis of the variation of $\eta$ as a function of 
$\mu_{\rm lim}$, as a possible diagnostic for an outer shallow exponential slope, in conjunction with the core-envelope color contrast technique by 
\citet[][cf. their Fig. 15]{PapOst12} that permits identification of systems surrounded by an extended nebular envelope. 
Indeed, the recent realization that such systems are virtually omnipresent at high $z$'s \citep[e.g.,][]{Wisotzki18}, as conjectured by \citet{PapOst12}, calls for an exploration of suitable approaches for their structural and morphological characterization. 

This also applies to the morphologically diverse population of generally irregular galaxies at high $z$.
For instance, a fraction of $\sim$10\% of these systems show cometary morphology \citep[also referred to as tadpoles;][see also F\"orster-Schreiber et al. 2011 for further examples at $z\sim2$]{Elm07,Straughn06} that is due to the confinement of intense SF activity at the one tip of an elongated stellar host or clumpy proto-disk. Interestingly, cometary morphology is rather common among extremely metal-poor (12 + log(O/H) $\leq$ 7.6) young dwarf galaxy candidates in the local universe \citep{P08,ML11}, which points to its association with early and intermediate stages of dwarf galaxy evolution.\\ 

1D surface photometry is obviously insufficient for a full characterization of such irregular elongated galaxies, since,
following the standard definition, a SBP records the photometric radius \rr\ of the circle subtending the same area ${\cal A}$ 
as the isophote of a galaxy at each surface brightness level $\mu$.
\begin{figure*}[h]
\centering
\makebox[\textwidth][c]{\includegraphics[width=1.0\linewidth]{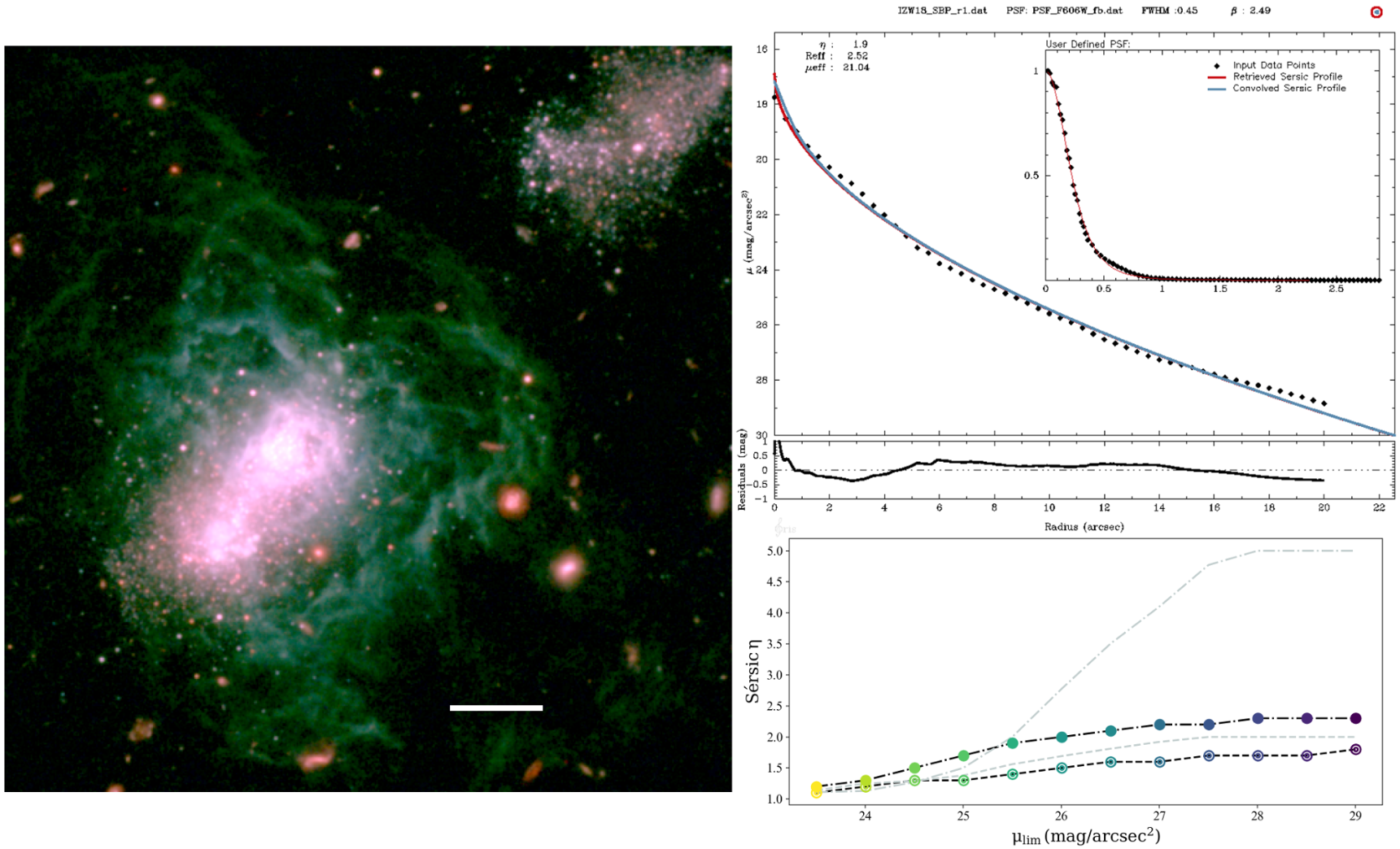}}
\caption[]{\brem{Left panel:} True-color image composite of \object{I Zw 18} produced from archival HST data, showing the extended nebular halo of the BCD \citep[adapted from][]{PapOst12}. The horizontal bar corresponds to 5\arcsec.
\brem{Upper-right panel:} \SL\ model with \ifit\ (blue) to the observed $R$-band SBP of \object{I Zw 18} \citep[SBP1; from][]{PapOst12}. The inset shows a Moffat model of the PSF (see Appendix for further explanations) and the fit residuals are displayed in the lower diagram. \brem{Bottom panel:} Variation of the best-fitting S\'ersic index $\eta$ vs. limiting surface brightness \mulim\ obtained by \citet{PapOst12} for SBP1 and SBP2 using standard $\chi^2$ minimization (gray dashed and dashed-dotted curve, respectively). Open and filled circles show the result obtained with \ifit\ by modeling both SBPs down to a \mulim\ between 23.5 \sbb\ and 29 \sbb.}\label{IZw18}
\end{figure*}
This sacrifices important morphological information (e.g., the variation of ellipticity $\epsilon$, position angle PA and $\alpha_4$ across $\mu$), unless the SBP determination is supplemented by, e.g., a Fourier analysis of the best-fitting ellipses to galaxy isophotes \citep[e.g.,][]{Carter1978,LooseThuan1986b,Bender1988,Caon93}.
In the case of cometary/tadpole galaxies and other irregular systems that strongly depart from axis-symmetry, the determination of ${\cal A}$ via fitting of ellipses to isophotes (or extraction of an SBP within concentric elliptical apertures with fixed center, PA and $\epsilon$) is non-optimal. Preferable SBP extraction 
techniques in those cases involve the computation of ${\cal A}$ through a line integral along isophotes or summation of pixels above an intensity threshold \citep[cf. methods ii \& iii in P96a, see also][]{LooseThuan1986b}. A probably even better approach is to compute photon statistics within irregular isophotal annuli (\emph{isan}) 
that closely trace the morphology of a galaxy at each $\mu$-intervall \citep[][see also Noeske et al. 2002, 2005, 2006 for its 
application to local and higher-$z$ star-forming galaxies]{P02}. This technique, eventually supplemented by 
a Fourier analysis of \emph{isan} jointly with quantitative morphology indicators (e.g., asymmetry index) might offer a promising avenue to the characterization 
cometary and irregular galaxies near and far. Such an approach, combined with \ifit, would probably be preferable to currently existing parametric 2D photometry tools since all latter are essentially tied to the simplifying assumption that galaxies can be approximated by a superposition of axis-symmetric components. It should also be noted that most structural quantities inferred from 2D photometry and considered in galaxy studies are actually converted into 1D quantities. For instance, one finally adopts one single value for the S\'ersic index $\eta$ instead of its best-fitting value along the major and minor axis light profile, although these two can substantially differ from one another \citep[e.g.,][]{Ferrari04}. The same applies to the model-dependent \reff, which despite its intrinsic coupling with $\eta$ is widely used as a measure of the "size" of a galaxy.

In the light of the above considerations it appears worthwhile to further explore quantitative morphological and structural indicators for irregular cometary/tadpole
galaxies. Devising methods for the identification and characterization of such elongated systems near and far appears to be timely given that they may have an appreciable impact on the weak lensing signal \citep{Pandya19} with, for instance, Euclid.

\section{Summary and conclusions \label{conc}}
In this article, we present \ifit, a simple yet robust algorithm for fitting a S\'ersic model of galaxy SBPs. 
At variance to standard \SL\ fitting routines that employ direct $\chi^2$ minimization between fit and individual SBP data points (or image pixels), the distinctive characteristic of \ifit\ is that it identifies the best-fitting \SL\ model of a SBP through the search for the best match between the observationally determined and theoretically expected light growth curve and radial variation of the mean surface brightness.
This novel concept therefore effectively ties the fit solution to robust photometric quantities (instead of individual SBP data points that can be subject to large uncertainties) and warrants quick convergence to a unique solution, even for shallow and resolution-degraded SBPs of galaxies showing significant deviations from the \SL.
Indeed, a series of tests on synthetic and observed SBPs indicate that solutions with \ifit\ generally show little dependence both on limiting surface brightness, quality of sky subtraction and PSF convolution effects, as well as on the photometric band considered. 
The reliability of the code is further underscored by the fact that subtraction of the \SL\ models independently obtained with \ifit\ for local early-type galaxies in two different filters closely reproduces the observed 
radial color profile. Furthermore, the capability of \ifit\ to robustly infer the best-fitting equivalent S\'ersic model for irregular galaxies that strongly deviate from the \SL\ (as most high-$z$ galaxies) is demonstrated  
on the analysis of the local BCDs \object{He 2-10} and \object{I Zw 18}.
For these reasons, and given that \ifit\ does not require an initial guess to \SL\ model parameters, it offers a robust and computationally 
inexpensive tool for the automated structural analysis of galaxies near and far.

\begin{acknowledgements}
We would like to thank the anonymous referee for numerous valuable comments and suggestions. 
We thank the European taxpayer, who in the spirit of solidarity between EU countries has offered
to Portugal a substantial fraction of the financial resources that allowed it to sustain a research infrastructure in astrophysics. Specifically, the major part of this work was carried out at an institute whose funding is provided to $\sim$85\% by the EU via the FCT (Funda\c{c}\~{a}o para a Ci\^{e}ncia e a Tecnologia) apparatus, through European and national funding via FEDER through COMPETE by the grants UID/FIS/04434/2013 \& POCI-01-0145-FEDER-007672 and PTDC/FIS-AST/3214/2012 \& FCOMP-01-0124-FEDER-029170. 
Additionally, this work was supported by FCT/MCTES through national funds (PIDDAC) by this grant UID/FIS/04434/2019.

We acknowledge support by European Community Programme (FP7/2007-2013) under grant agreement No. PIRSES-GA-2013-612701 (SELGIFS). 
I.B. was supported by the FCT PhD::SPACE Doctoral Network (PD/00040/2012) through the fellowship PD/BD/52707/2014 funded by FCT (Portugal) and POPH/FSE (EC) and by the fellowship CAUP-07/2014-BI in the context of the FCT project PTDC/FIS-AST/3214/2012 \& FCOMP-01-0124-FEDER-029170, as well as by the fellowship CIAAUP-19/2019-BIM within the scope of the research unit Instituto de Astrof\'{i}sica e Ci\^{e}ncias do Espaço (IA).
P.P. was supported through Investigador FCT contract IF/01220/2013/CP1191/CT0002 and by a contract 
that is supported by FCT/MCTES through national funds (PIDDAC) and by grant PTDC/FIS-AST/29245/2017.
J.M.G. is supported by the fellowship CIAAUP-04/2016-BPD in the context of the FCT project UID/FIS/04434/2013 \& POCI-01-0145-FEDER-007672 and acknowledges the previous support by the fellowships SFRH/BPD/66958/2009 funded by FCT and POPH/FSE (EC) and DL 57/2016/CP1364/CT0003.
SA gratefully acknowledge support from the Science and Technology Foundation (FCT, Portugal) through the research grants PTDC/FIS-AST/29245/2017 and UID/FIS/04434/2019.
We thank Sandra dos Reis for granting us access to her results with \galfit\ prior to publication.
This research has made use of the NASA/IPAC Extragalactic Database (NED) which is operated by the Jet Propulsion Laboratory, 
California Institute of Technology, under contract with the National Aeronautics and Space Administration.
\end{acknowledgements}


\onecolumn 
\begin{appendices}
\section{Testing iFIT on real galaxies}\label{appA}

\ifit\ was tested on SDSS $g$ \& $r$-band data for 121 local ETGs. The images were sky-subtracted, rotated in astronomical orientation and corrected for Galactic foreground extinction \citep[adopting the values by][available at NASA Extragalactic Data Base--NED]{SchFin11}.
Galactic foreground stars that overlap some ETGs were substituted by the mean intensity of the adjacent stellar continuum. Finally the frames were trimmed and the SBPs were extracted using the function FIT/ELL3 within the package SURFPHOT of ESO-MIDAS. 
Below we show only results from fitting in the $g$-band, whereby the effect of the adopted PSF was examined by running \ifit\ three times for each galaxy using a different non-saturated field star to approximate the PSF. 
Although some PSFs render better results than others at small radii, it can be seen from Fig. \ref{fig1A} and Table~\ref{tabA} that the result from \ifit\ does not appreciably depend on PSF corrections, except for the cases when \minpack\ is unable to properly fit the PSF. 

\captionsetup{width=18cm}
\setlength{\LTleft}{2.8cm}
\begin{small}
\begin{longtable}{ccccccccccc}
\label{tabA}
Name & $\eta_{1}$ & R$_{\rm eff \, 1}$ & $\mu_{\rm eff \, 1}$ & $\eta_{2}$ & R$_{\rm eff \, 2}$ & $\mu_{\rm eff \, 2}$ & $\eta_{3}$ & R$_{\rm eff \, 3}$ & $\mu_{\rm eff \, 3}$ & Time (s) \\ 
IC1079 & 3.70 & 34.33 & 24.10 & 3.70 & 34.26 & 24.09 & 3.70 & 34.39 & 24.10 & 1.62 \\
IC2341 & 2.20 & 9.87 & 21.74 & 2.20 & 9.87 & 21.74 & 2.20 & 9.49 & 21.54 & 0.69 \\
IC4534 & 2.30 & 12.01 & 21.82 & 2.70 & 17.73 & 22.65 & 2.70 & 17.80 & 22.66 & 1.03 \\
LSBCF560-04 & 2.10 & 15.09 & 22.83 & 2.10 & 15.10 & 22.83 & 2.10 & 15.07 & 22.83 & 2.35 \\
NGC0155 & 2.90 & 19.17 & 22.69 & 2.70 & 17.43 & 22.47 & 2.70 & 17.41 & 22.47 & 1.08 \\
NGC0160 & 4.00 & 48.65 & 24.24 & 4.00 & 48.42 & 24.23 & 4.00 & 48.42 & 24.23 & 1.37 \\
NGC0364 & 3.10 & 15.69 & 22.68 & 3.10 & 15.78 & 22.70 & 3.10 & 15.75 & 22.69 & 1.19 \\
NGC0472 & 3.50 & 13.46 & 22.53 & 3.50 & 13.47 & 22.54 & 3.50 & 13.16 & 22.52 & 2.2 \\
NGC0499 & 2.50 & 16.85 & 21.58 & 2.60 & 17.26 & 21.63 & 2.60 & 17.40 & 21.65 & 3.75 \\
NGC0517 & 2.90 & 10.57 & 21.53 & 2.90 & 10.57 & 21.53 & 2.90 & 10.57 & 21.53 & 1.87 \\
NGC0528 & 2.60 & 10.92 & 21.42 & 2.60 & 10.92 & 21.42 & 2.60 & 10.93 & 21.42 & 1.75 \\
NGC0529 & 3.30 & 18.14 & 22.08 & 2.90 & 17.35 & 21.94 & 2.80 & 15.62 & 21.74 & 1.84 \\
NGC0677 & 3.10 & 17.29 & 22.17 & 3.10 & 17.87 & 22.24 & 3.00 & 16.91 & 22.11 & 1.72 \\
NGC0681 & 2.10 & 26.18 & 22.28 & 2.40 & 31.88 & 22.70 & 2.10 & 26.24 & 22.29 & 3.49 \\
NGC0731 & 2.70 & 15.14 & 21.65 & 2.50 & 15.41 & 21.66 & 2.50 & 15.38 & 21.65 & 1.51 \\
NGC0741 & 2.60 & 29.85 & 22.36 & 2.60 & 29.87 & 22.36 & 2.60 & 29.86 & 22.36 & 3.14 \\
NGC0774 & 2.20 & 12.38 & 21.76 & 2.10 & 12.55 & 21.78 & 2.10 & 12.56 & 21.78 & 1.23 \\
NGC0787 & 3.60 & 41.10 & 23.70 & 3.60 & 37.60 & 23.54 & 3.00 & 25.96 & 22.91 & 1.55 \\
NGC0810 & 2.30 & 18.00 & 22.33 & 2.30 & 18.00 & 22.33 & 2.30 & 18.01 & 22.33 & 1.35 \\
NGC0842 & 3.20 & 15.62 & 22.11 & 3.00 & 15.39 & 22.07 & 2.90 & 14.10 & 21.88 & 1.06 \\
NGC0924 & 3.30 & 16.69 & 21.64 & 3.30 & 16.69 & 21.64 & 3.30 & 16.69 & 21.64 & 2.23 \\
NGC0932 & 3.60 & 32.45 & 22.82 & 3.70 & 35.84 & 22.98 & 3.60 & 32.30 & 22.81 & 2.41 \\
NGC0938 & 3.30 & 15.54 & 21.79 & 3.60 & 19.09 & 22.18 & 3.30 & 15.55 & 21.79 & 3.05 \\
NGC0962 & 2.50 & 14.92 & 21.63 & 2.50 & 14.93 & 21.63 & 2.50 & 14.93 & 21.63 & 1.23 \\
NGC1026 & 3.30 & 25.65 & 22.69 & 2.90 & 22.79 & 22.45 & 3.00 & 24.43 & 22.57 & 1.43 \\
NGC1041 & 4.00 & 26.14 & 23.52 & 3.00 & 18.96 & 22.82 & 3.10 & 20.65 & 22.98 & 1.46 \\
NGC1056 & 2.30 & 14.90 & 21.24 & 2.30 & 14.52 & 21.18 & 2.40 & 14.98 & 21.26 & 0.88 \\
NGC1060 & 2.70 & 24.63 & 21.36 & 2.70 & 25.71 & 21.43 & 2.70 & 24.63 & 21.36 & 1.25 \\
NGC1132 & 3.20 & 36.83 & 23.51 & 2.60 & 26.48 & 22.86 & 2.60 & 26.48 & 22.86 & 1.12 \\
NGC1167 & 2.80 & 31.61 & 22.21 & 2.80 & 31.61 & 22.21 & 3.10 & 38.00 & 22.57 & 1.14 \\
NGC1270 & 2.70 & 9.34 & 20.80 & 2.90 & 10.01 & 20.97 & 2.70 & 9.35 & 20.80 & 1.09 \\
NGC1349 & 3.00 & 26.90 & 22.67 & 3.00 & 26.79 & 22.66 & 3.00 & 26.44 & 22.64 & 0.96 \\
NGC1361 & 3.20 & 18.54 & 23.10 & 3.10 & 16.50 & 22.88 & 3.10 & 16.47 & 22.88 & 1.41 \\
NGC1656 & 3.30 & 19.04 & 22.35 & 3.30 & 18.92 & 22.34 & 3.30 & 19.04 & 22.35 & 2.25 \\
NGC1665 & 4.00 & 42.95 & 23.68 & 4.00 & 42.68 & 23.67 & 4.00 & 42.68 & 23.67 & 1.83 \\
NGC1666 & 3.50 & 18.62 & 22.09 & 4.20 & 28.80 & 22.95 & 3.40 & 18.55 & 22.08 & 1.52 \\
NGC2476 & 3.20 & 12.73 & 21.62 & 3.20 & 12.75 & 21.63 & 3.20 & 12.73 & 21.62 & 1.49 \\
NGC2507 & 2.60 & 27.81 & 22.64 & 2.60 & 27.81 & 22.64 & 2.60 & 27.84 & 22.64 & 0.89 \\
NGC2513 & 3.70 & 38.83 & 23.29 & 3.00 & 27.91 & 22.64 & 3.00 & 27.49 & 22.61 & 1.42 \\
NGC2577 & 2.50 & 14.63 & 21.43 & 2.50 & 13.99 & 21.34 & 2.50 & 14.63 & 21.43 & 1.59 \\
NGC2592 & 3.30 & 13.58 & 21.46 & 2.70 & 11.22 & 21.04 & 2.70 & 11.22 & 21.04 & 1.41 \\
NGC2639 & 2.00 & 15.53 & 21.03 & 2.00 & 15.53 & 21.03 & 2.00 & 15.51 & 21.03 & 2.71 \\
NGC2767 & 2.90 & 11.53 & 22.42 & 2.90 & 11.26 & 22.37 & 2.90 & 11.33 & 22.38 & 0.83 \\
NGC2918 & 2.10 & 12.95 & 21.79 & 2.10 & 12.92 & 21.78 & 2.10 & 12.92 & 21.78 & 1.01 \\
NGC3106 & 4.90 & 68.69 & 24.96 & 3.80 & 32.68 & 23.60 & 3.80 & 32.70 & 23.60 & 2.75 \\
NGC3158 & 2.80 & 24.75 & 22.63 & 2.80 & 24.41 & 22.60 & 2.80 & 24.75 & 22.63 & 1.17 \\
NGC3182 & 2.80 & 19.79 & 22.28 & 2.70 & 19.25 & 22.22 & 2.70 & 19.11 & 22.21 & 2.23 \\
NGC3300 & 2.30 & 15.64 & 21.74 & 2.30 & 15.64 & 21.74 & 2.30 & 15.74 & 21.76 & 1.03 \\
NGC3610 & 2.70 & 15.06 & 20.29 & 2.70 & 15.06 & 20.29 & 2.70 & 15.06 & 20.29 & 2.15 \\
NGC3615 & 2.90 & 14.33 & 22.06 & 3.00 & 15.92 & 22.24 & 3.00 & 15.93 & 22.24 & 1.45 \\
NGC3619 & 3.90 & 47.05 & 23.46 & 3.60 & 38.57 & 23.07 & 3.80 & 43.84 & 23.33 & 2.66 \\
NGC3990 & 2.40 & 9.44 & 20.93 & 2.80 & 10.45 & 21.19 & 2.70 & 10.13 & 21.15 & 0.91 \\
NGC4003 & 2.30 & 13.53 & 22.49 & 2.30 & 13.61 & 22.51 & 2.30 & 13.57 & 22.50 & 0.96 \\
NGC4816 & 3.70 & 32.13 & 23.94 & 3.70 & 32.06 & 23.94 & 3.70 & 32.13 & 23.94 & 1.87 \\
NGC4841A & 3.10 & 26.63 & 23.23 & 3.10 & 26.49 & 23.22 & 3.10 & 26.63 & 23.23 & 1.75 \\
NGC4874 & 3.20 & 69.98 & 24.49 & 2.40 & 41.60 & 23.52 & 2.50 & 44.95 & 23.67 & 0.61 \\
NGC4956 & 2.40 & 12.11 & 21.33 & 2.40 & 11.84 & 21.29 & 2.40 & 11.78 & 21.28 & 1.69 \\
NGC5029 & 3.10 & 20.36 & 23.07 & 3.10 & 20.38 & 23.07 & 3.30 & 23.17 & 23.31 & 2.02 \\
NGC5198 & 2.90 & 24.52 & 22.30 & 2.90 & 24.82 & 22.32 & 2.90 & 24.60 & 22.30 & 1.94 \\
NGC5216 & 3.00 & 25.42 & 23.25 & 3.00 & 25.56 & 23.26 & 3.00 & 25.42 & 23.25 & 1.27 \\
NGC5423 & 3.20 & 15.54 & 22.44 & 3.10 & 14.71 & 22.32 & 3.20 & 15.72 & 22.46 & 1.55 \\
NGC5473 & 4.00 & 37.11 & 22.87 & 2.70 & 19.47 & 21.55 & 3.20 & 23.25 & 21.96 & 1.74 \\
NGC5481 & 2.70 & 33.29 & 22.94 & 2.70 & 33.04 & 22.93 & 2.70 & 33.29 & 22.94 & 1.28 \\
NGC5485 & 2.30 & 25.42 & 21.99 & 2.30 & 25.39 & 21.99 & 2.40 & 25.12 & 21.98 & 1.49 \\
NGC5513 & 3.20 & 20.82 & 22.49 & 3.40 & 18.17 & 22.13 & 3.00 & 18.55 & 22.25 & 1.25 \\
NGC5525 & 2.40 & 16.64 & 22.31 & 2.40 & 16.64 & 22.31 & 2.40 & 16.62 & 22.30 & 1.4 \\
NGC5532 & 2.60 & 18.59 & 18.04 & 2.60 & 19.17 & 18.10 & 2.60 & 18.72 & 18.06 & 3.25 \\
NGC5546 & 3.20 & 20.84 & 22.69 & 3.00 & 20.77 & 22.67 & 3.00 & 20.79 & 22.67 & 1.57 \\
NGC5549 & 2.70 & 14.51 & 22.04 & 2.70 & 14.52 & 22.04 & 2.70 & 14.53 & 22.05 & 1.2 \\
NGC5557 & 3.50 & 30.90 & 22.27 & 3.50 & 30.90 & 22.27 & 2.70 & 19.88 & 21.34 & 3.18 \\
NGC5580 & 2.30 & 15.60 & 22.01 & 2.50 & 17.07 & 22.20 & 2.40 & 16.88 & 22.16 & 1.56 \\
NGC5598 & 2.50 & 10.17 & 21.74 & 2.50 & 10.18 & 21.74 & 2.50 & 10.34 & 21.79 & 1.55 \\
NGC5611 & 2.40 & 7.68 & 20.67 & 2.40 & 7.72 & 20.68 & 2.40 & 7.64 & 20.66 & 1.38 \\
NGC5614 & 2.80 & 19.80 & 21.80 & 2.80 & 19.76 & 21.79 & 2.80 & 19.76 & 21.79 & 1.8 \\
NGC5623 & 3.50 & 17.97 & 22.47 & 3.50 & 17.97 & 22.47 & 3.50 & 17.94 & 22.47 & 2.12 \\
NGC5631 & 3.40 & 24.83 & 22.02 & 3.40 & 24.66 & 22.01 & 3.40 & 24.66 & 22.01 & 3.04 \\
NGC5642 & 3.60 & 18.72 & 22.38 & 3.20 & 18.25 & 22.30 & 3.20 & 18.25 & 22.30 & 2.35 \\
NGC5684 & 3.00 & 20.40 & 22.89 & 3.10 & 20.93 & 22.95 & 2.90 & 19.30 & 22.75 & 1.43 \\
NGC5687 & 3.20 & 24.19 & 22.53 & 3.10 & 23.43 & 22.46 & 3.20 & 24.41 & 22.55 & 1.65 \\
NGC5784 & 2.60 & 15.72 & 21.93 & 2.60 & 15.73 & 21.94 & 2.60 & 15.74 & 21.94 & 1.7 \\
NGC5797 & 2.90 & 12.62 & 21.55 & 3.00 & 13.25 & 21.66 & 2.90 & 12.62 & 21.55 & 1.26 \\
NGC5928 & 3.10 & 18.42 & 22.29 & 3.10 & 18.50 & 22.30 & 3.10 & 18.47 & 22.30 & 1.9 \\
NGC5966 & 2.70 & 16.67 & 22.30 & 2.70 & 16.09 & 22.23 & 2.70 & 16.67 & 22.30 & 1.44 \\
NGC6020 & 3.40 & 16.87 & 22.31 & 3.20 & 16.91 & 22.30 & 3.50 & 16.98 & 22.33 & 1.46 \\
NGC6021 & 3.30 & 15.45 & 22.49 & 2.60 & 12.14 & 21.94 & 2.60 & 12.53 & 22.01 & 1.46 \\
NGC6023 & 3.00 & 18.42 & 23.05 & 3.20 & 20.81 & 23.29 & 3.00 & 18.48 & 23.06 & 1.51 \\
NGC6081 & 2.40 & 14.48 & 22.24 & 2.40 & 14.74 & 22.29 & 2.40 & 14.70 & 22.28 & 1.04 \\
NGC6125 & 3.10 & 19.52 & 22.27 & 2.70 & 17.88 & 22.09 & 2.70 & 17.97 & 22.10 & 2.95 \\
NGC6146 & 3.00 & 16.28 & 22.35 & 2.80 & 14.42 & 22.10 & 2.90 & 14.94 & 22.18 & 1.3 \\
NGC6150 & 2.20 & 10.50 & 22.03 & 2.00 & 10.06 & 21.92 & 2.00 & 10.06 & 21.92 & 0.66 \\
NGC6166NED01 & 1.60 & 25.73 & 22.74 & 1.60 & 25.71 & 22.74 & 1.60 & 25.68 & 22.73 & 0.4 \\
NGC6173 & 3.20 & 23.29 & 22.80 & 3.20 & 23.28 & 22.80 & 3.20 & 23.29 & 22.80 & 2.22 \\
NGC6278 & 4.00 & 24.12 & 22.60 & 3.00 & 13.46 & 21.41 & 3.00 & 13.46 & 21.41 & 1.59 \\
NGC6338 & 3.20 & 28.93 & 23.30 & 3.20 & 29.22 & 23.32 & 3.20 & 29.32 & 23.33 & 1.6 \\
NGC6411 & 3.80 & 24.80 & 22.33 & 3.30 & 26.19 & 22.44 & 3.10 & 23.57 & 22.24 & 3.3 \\
NGC6515 & 3.50 & 20.82 & 22.90 & 3.60 & 22.13 & 23.02 & 3.50 & 20.82 & 22.90 & 1.82 \\
NGC6762 & 1.90 & 8.77 & 21.22 & 2.00 & 8.54 & 21.16 & 1.90 & 8.77 & 21.22 & 0.69 \\
NGC7025 & 2.90 & 20.51 & 21.99 & 2.90 & 20.51 & 21.99 & 3.30 & 25.33 & 22.43 & 1.82 \\
NGC7194 & 2.90 & 13.86 & 22.17 & 2.80 & 13.53 & 22.12 & 2.80 & 13.16 & 22.06 & 1.08 \\
NGC7236 & 2.70 & 32.67 & 23.63 & 2.70 & 32.63 & 23.63 & 2.70 & 32.67 & 23.63 & 1.09 \\
NGC7550 & 2.80 & 24.65 & 22.05 & 2.80 & 24.65 & 22.05 & 2.60 & 21.68 & 21.77 & 1.47 \\
NGC7559B & 3.40 & 27.95 & 23.36 & 2.90 & 19.08 & 22.69 & 3.40 & 28.02 & 23.37 & 1.54 \\
NGC7611 & 3.30 & 13.26 & 21.50 & 2.90 & 10.72 & 21.01 & 2.90 & 10.76 & 21.02 & 1.54 \\
NGC7619 & 3.10 & 25.49 & 21.68 & 3.10 & 25.49 & 21.68 & 3.10 & 25.49 & 21.68 & 1.55 \\
NGC7671 & 2.90 & 12.38 & 21.45 & 2.90 & 12.40 & 21.46 & 2.90 & 12.36 & 21.45 & 1.51 \\
NGC7683 & 2.70 & 16.12 & 21.77 & 2.70 & 16.08 & 21.77 & 2.70 & 16.13 & 21.78 & 1.44 \\
NGC7711 & 3.10 & 17.17 & 21.83 & 3.00 & 15.83 & 21.65 & 3.00 & 15.83 & 21.65 & 1.63 \\
NGC7722 & 3.20 & 36.35 & 23.39 & 3.40 & 37.89 & 23.49 & 2.90 & 30.43 & 23.04 & 1.22 \\
UGC00029 & 2.60 & 14.75 & 22.88 & 2.60 & 14.76 & 22.88 & 2.60 & 14.75 & 22.88 & 1.04 \\
UGC03960 & 3.00 & 18.79 & 23.18 & 3.00 & 18.78 & 23.18 & 3.00 & 18.76 & 23.18 & 1.09 \\
UGC05771 & 3.30 & 15.38 & 22.99 & 3.30 & 15.39 & 23.00 & 3.30 & 15.34 & 22.99 & 1.47 \\
UGC08234 & 2.80 & 9.24 & 21.37 & 2.90 & 9.44 & 21.41 & 2.80 & 9.23 & 21.36 & 1.27 \\
UGC09518 & 3.40 & 13.05 & 22.47 & 3.40 & 13.05 & 22.47 & 3.90 & 15.03 & 22.80 & 2.2 \\
UGC10097 & 3.20 & 17.27 & 22.48 & 3.20 & 17.27 & 22.48 & 3.40 & 16.77 & 22.43 & 1.74 \\
UGC10205 & 1.70 & 20.16 & 22.85 & 1.70 & 19.68 & 22.79 & 1.60 & 19.05 & 22.73 & 0.57 \\
UGC10693 & 2.80 & 16.44 & 22.36 & 2.80 & 16.44 & 22.36 & 2.80 & 16.45 & 22.36 & 1.66 \\
UGC10695 & 2.90 & 20.03 & 23.33 & 2.90 & 20.33 & 23.36 & 2.90 & 20.18 & 23.35 & 1.09 \\
UGC10905 & 3.40 & 15.09 & 22.38 & 3.40 & 15.10 & 22.38 & 3.40 & 15.09 & 22.38 & 1.84 \\
UGC11228 & 2.80 & 12.44 & 22.07 & 2.80 & 12.44 & 22.07 & 2.80 & 12.44 & 22.07 & 0.95 \\
UGC11958 & 6.70 & 518.32 & 28.27 & 4.10 & 44.49 & 24.27 & 4.10 & 44.55 & 24.27 & 2.24 \\
UGC12127 & 2.60 & 20.62 & 22.69 & 2.60 & 21.14 & 22.71 & 2.60 & 21.18 & 22.72 & 0.8 \\
\caption{\SL\ model parameters obtained with \ifit\ by fitting $g$-band SDSS images for three times each
using different PSFs that were computed from nearby field stars.
The table lists, from the left to the right: name of the galaxy, and the derived $\eta$, \reff\ and $\mu_{\rm eff}$ for the three different
fits. The last column tabulates the average CPU expense in seconds on a laptop equipped with an Intel i7-2620M CPU @ 2.70GHz. 
}
\end{longtable}
\end{small}

\newpage

The following figure shows the output from \ifit\ for the ETG NGC4956 for two PSF models:

\captionsetup{width=18cm}
\begin{figure}[h]
\makebox[\textwidth][c]{\includegraphics[width=1.0\linewidth]{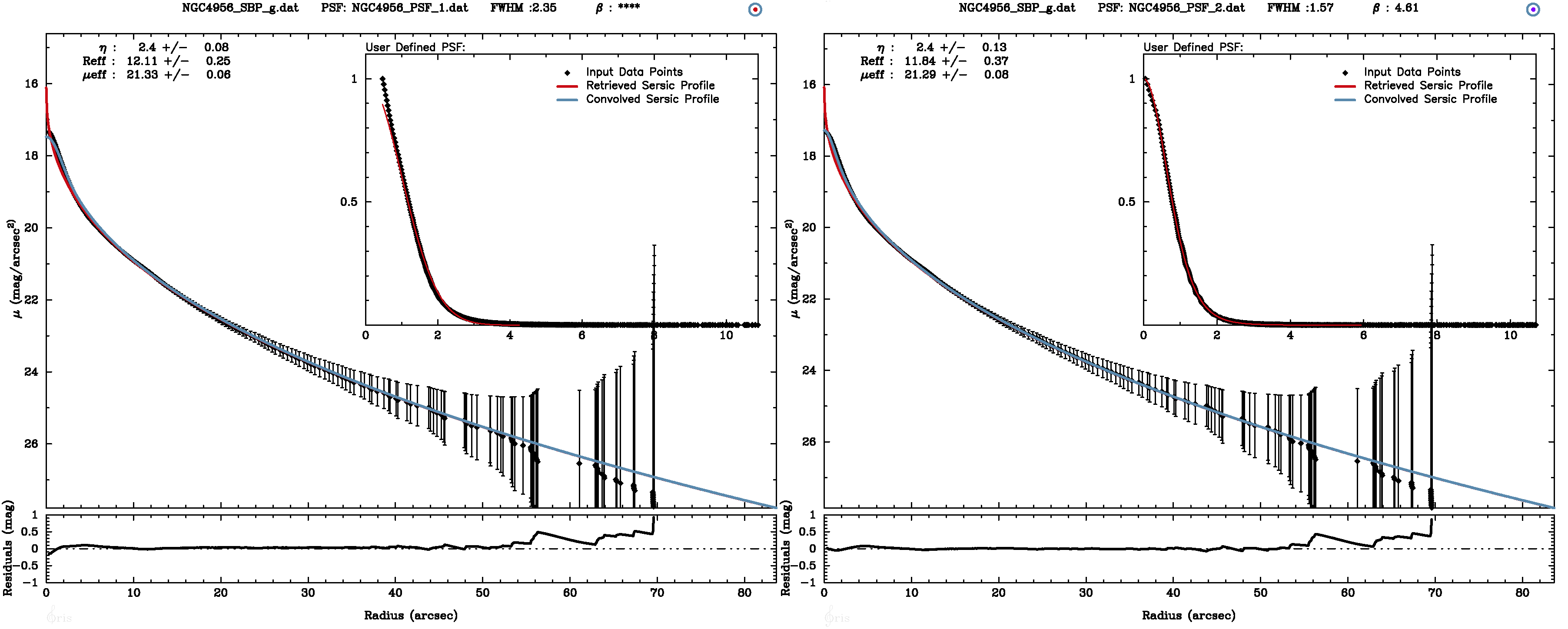}}
\caption{Fitting of the SBP of NGC4956 with \ifit\ using two different PSFs. The SBPs (both from the galaxy and individual stars used for the PSF determination) were obtained with the routine FIT/ELL3 in ESO-MIDAS for $g$-band frames from SDSS DR7. The graphical output from \ifit\ provides in the header the following information: name of the input file, name of the PSF file, and estimated characteristics for the input PSF (FWHM and $\beta_{\rm M}$). 
The color of the inner dot in the r.h.s. circle informs on the method used to estimate \reff\ and $\mu_{\rm eff}$: blue -- non-weighted linear regression; purple -- $w = 1/\sqrt{\sigma_{\mu}}$; orange -- $w = 1/\sigma_{\mu}$; red -- $w = 1/\sigma_{\mu}^2$ and that of the 
circle gives an estimate of the goodness of the fit with blue, purple and red corresponding to an adequate, modest quality and poor fit.
Main plot: black dots show the observed SBP and its uncertainties. The red and blue curve shows, respectively, the best-fitting \SL\ model and its convolution with the PSF, and the top-left labels list the estimated $\eta$, \reff\ (\arcsec) and $\mu_{\rm eff}$ (\sbb). The inset shows the user-defined PSF (i.e., the radial profile of one or several nearby non-saturated field stars) and the fit to it with \minpack\ routines.
A $\beta$ = **** indicates that \minpack\ failed to correctly fit the PSF, as it can be seen from inspection of the inset in the l.h.s. plot. The bottom panel displays the residuals (in mag) between the observed SBP and the best-fitting \SL\ model of it.  
}\label{fig1A}
\end{figure} 

Additionally \ifit\ was tested on F435W and F606W from ACS/WFC imaging data for higher-z galaxies extracted from the Hubble eXtreme Deep Field (XDF) Data release 1.0 \citep{Ill13}. Figure \ref{fig2A} shows two examples of such galaxies.

\captionsetup{width=18cm}
\begin{figure}[h]
\makebox[\textwidth][c]{\includegraphics[width=1.0\linewidth]{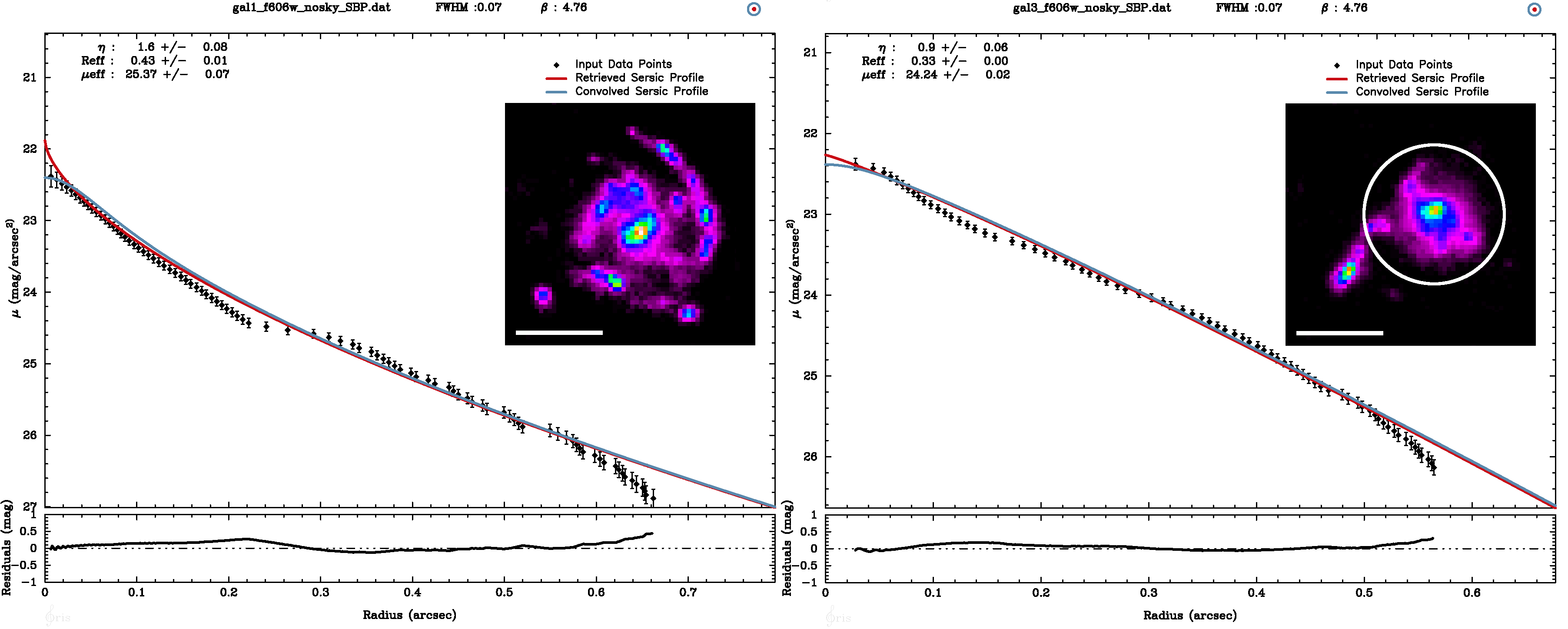}}
\caption{Fits with \ifit\ of the SBP of two higher-$z$ galaxies from XDF in the F606W filter after background subtraction. The upper and lower panel displays, respectively, the galaxy ACS-GC 90045619 (RA: 53.157796$^{\circ}$, DEC: -27.797536$^{\circ}$) at $z$ = 0.768 \citep{Bond14} and 
UDF:[JBM2015]\,18620 (RA: 53.167567$^{\circ}$, DEC: -27.792520$^{\circ}$) at $z$ = 1.091 \citep{Jim15} (encircled). The horizontal bar corresponds to 1\arcsec.}   
\label{fig2A}
\end{figure}

\clearpage


\section{Empirical assessment of the alternative weights in the weighted linear regression}\label{appB}
Weighted least squares (WLS) (a variation of the ordinary least squares, OLS) should be applied when the assumption of constant variance in the errors is violated (heteroscedasticity), the measurements come from a Gaussian distribution and the stochastic variation from point to point is independent. 

Providing that there is a functional relation ($f$) between the observational points ($y$) such as $y = f(x,a,b)$, being $x$ and $y$ the independent and dependent variables, respectively, and $a$ and $b$ the model parameters to be determined, WLS minimizes the $\chi^2$ statistics by varying the parameters $a$ and $b$:

\vspace*{-0.5cm}
\begin{equation}
\chi^2 = \sum\limits_{i=1}^{n} w(i) \cdot (y_i - f(x_i,a,b))^2,
\label{eqB1}
\end{equation}

where the optimal weight to be adopted is the reciprocal of the variance of the measured errors -- $w(i) = 1/\sigma_{\mu}(i)^2$ \citep{Ati36}.

The aforementioned procedure is optimal in theory, that is, when the measured data points perfectly match the model to be fit. However, the essential message from this article is that galaxy SBPs do not necessarily obey the \SL, but in many cases show substantial systematic deviations from it, especially but not exclusively in their brightest central part due to, for example, depleted cores in massive ETGs, central nuclei in dwarf ellipticals, circumnuclear star-forming rings in LTGs (cf. Introduction). Using standard $\chi^2$-minimization algorithms to fit the \SL\ to such \iSPs\ unavoidably leads to solutions being primarily dictated by the central, hence lowest-$\sigma_{\mu}(i)$, SBP data points which often fail to adequately describe the extended lower-surface brightness component of galaxies, being unable to properly model the underlying \SL\ component. While such fits might be irreproachable in terms of their reduced $\chi^2$, they can result 
in substantial and possibly systematic errors in several widely used quantities for the structural characterization of galaxies 
(e.g., the model-dependent \reff\ as a measure of the size of a galaxy) and in the colors implied by \SL\ models for their LSB periphery.
The fact that these lowest-$\sigma_{\mu}(i)$ SBP data points (or image pixels) are precisely those mostly affected by PSF convolution 
effects further aggravates the problem. Although this drawback is rather obvious, it has actually not been addressed in any 1D/2D structural characterization of galaxies via the \SL, and we are unaware of a  quantitative assessment of the bias it could introduce in automated surface photometry studies of galaxies.

In the interest of overcoming this obstacle and developing a suitable approach for the determination of the best-fitting 
equivalent \SL\ model of an \iSP, we use here alternative weights, such as $w(i) = 1/\sqrt{\sigma_{\mu}(i)}$ and $w(i) = 1/\sigma_{\mu}(i)$, in order to decrease the relative impact of the innermost low-$\sigma_{\mu}(i)$ data points on the fitting solution. Below we attempt to empirically demonstrate the superiority of these alternative weights for recovering the true \SL\ parameters by fitting a few typical \iSP's. To avoid biases due to, for example, the lack of resolution or the shallowness of the input SBPs, we designed those such as to have a fine radius step (0\farcs01) and to contain $\sim$95\% of the total theoretical luminosity):

\captionsetup{width=18cm}
\begin{figure}[h]
\makebox[\textwidth][c]{\includegraphics[width=1.0\linewidth]{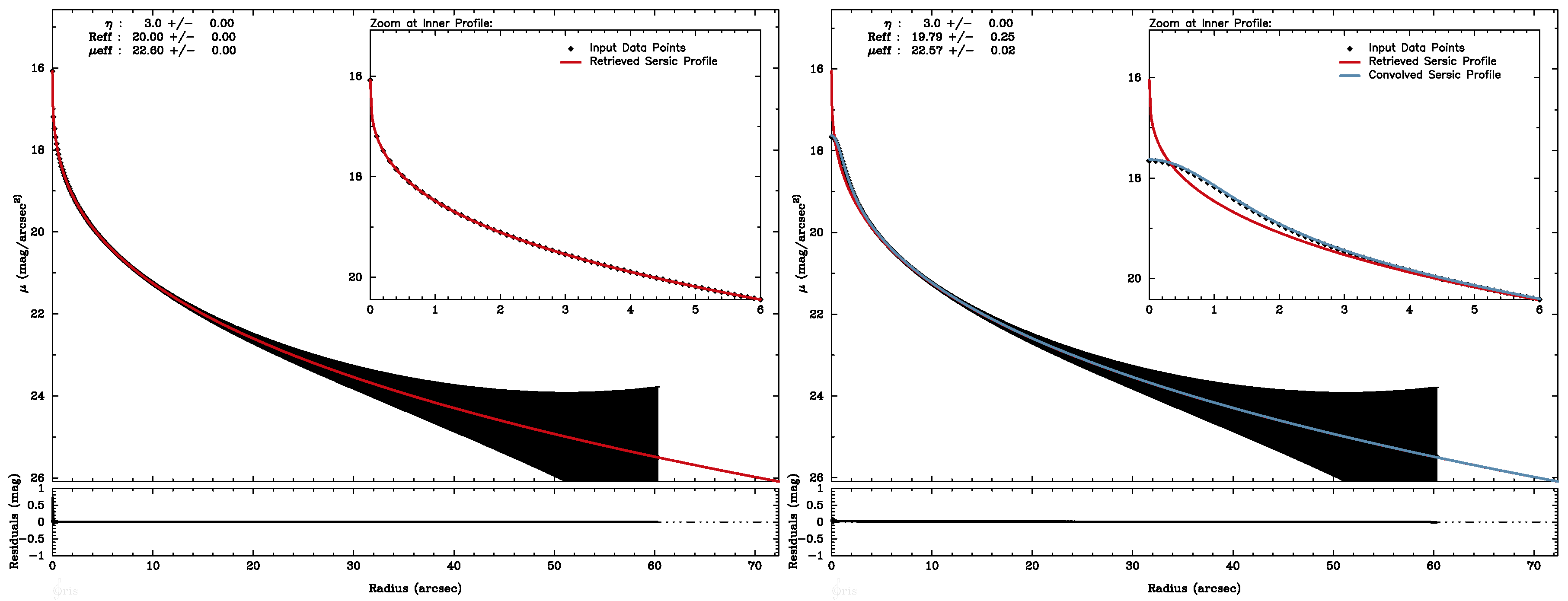}}
\caption[]{\brem{Left panel:} theoretical \SL\ profile with $\eta = 3$, \reff\ = 20\arcsec, $\mu_{\rm eff} = 22.6$ mag (hereafter \brem{pA}), with $\sigma_{\mu}$ estimated following \citet{P96a}, and its fit (red curve) with \ifit. \brem{Right panel:} Profile \brem{pA} after convolution with a Moffat PSF with FWHM of 1\farcs57 (hereafter, \SL\ profile \brem{pB}). The best-fitting \SL\ model and its convolution with the PSF are shown with the red and blue curve, respectively.}\label{Fig_AppB_1} 
\end{figure} 

\clearpage

\begin{multicols}{2}
1) The left panel of Fig.~\ref{Fig_AppB_1} illustrates the case of a perfect \SL\ profile ($\eta = 3$, \reff\ = 20\arcsec, $\mu_{\rm eff} = 22.6$ mag), denoted in the following as \brem{pA}. 
In the following tables, the middle column ($\eta_3$) lists the results obtained by performing linear regression via OLS/WLS where $x = \mathrm{R}^{1/\eta}$ with a fixed $\eta = 3$ and $y = \mu(\mathrm{R}^{\star})$, and subsequently applying Eg.~\ref{eq4a} (see Sect.~\ref{non-PSF}, item \textbf{3.}), adopting the different weights as specified in the leftmost column. The right-hand column lists the final solutions obtained by \ifit:

\begin{center}
\begin{tabular}{ c | p{0.3cm} c c | p{0.3cm} c c c}
 $w$ & \footnotesize $\eta_3$: & $\rm R_{eff}$ & $\mu_{\rm eff}$ & \footnotesize \ifit: & $\eta$ & $\rm R_{eff}$ & $\mu_{\rm eff}$ \\ 
 no weight & & 20.0 & 22.6 & & 3.0 & 20.0 & 22.6\\ 
 $1/\sqrt{\sigma_{\mu}}$ & & 20.0 & 22.6 & & 3.0 & 20.0 & 22.6\\ 
 $1/\sigma_{\mu}$ & & 20.0 & 22.6 & & 3.0 & 20.0 & 22.6\\ 
 $1/\sigma_{\mu}^2$ & & 20.0 & 22.6 & & 3.0 & 20.0 & 22.6\\ 
\end{tabular}
\end{center}

It is important to note that in the case of a perfect \SL\ profile, performing WLS with alternative weights returns the same solution as the one obtained by adopting the statistically optimal ($1/\sigma_{\mu}^2$) weights or OLS, confirming that the inclusion of the alternative weights wont introduce a bias.

\columnbreak

2) For the same \SL\ profile but convolved with a Moffat PSF with a FWHM of 1.57\arcsec\ (\brem{pB}, right panel of Fig.~\ref{Fig_AppB_1})
the same exercise yields:

\begin{center}
\begin{tabular}{ c | p{0.3cm} c c | p{0.3cm} c c c}
 $w$ & \footnotesize $\eta_3$: & $\rm R_{eff}$ & $\mu_{\rm eff}$ & \footnotesize \ifit: & $\eta$ & $\rm R_{eff}$ & $\mu_{\rm eff}$ \\ 
 no weight & & 19.8 & 22.6 & & 3.0 & 19.8 & 22.6\\ 
 $1/\sqrt{\sigma_{\mu}}$ & & 19.7 & 22.6 & & 3.0 & 19.7 & 22.6\\ 
 $1/\sigma_{\mu}$ & & 19.6 & 22.5 & & 3.0 & 19.6 & 22.5 \\ 
 $1/\sigma_{\mu}^2$ & & 19.2 & 22.5 & & 3.0 & 19.2 & 22.5 \\ 
\end{tabular}
\end{center}

These results show that even a slight alteration of the central low-$\sigma_{\mu}$ part of the profile (in this case, convolution with a Moffat function with FWHM/\reff\ $\approx$ 0.08) results in appreciable deviations from the correct solution (\reff\ = 20 and $\mu_{\rm eff}$ = 22.6) becoming larger when WLS fitting assumes $w = 1/\sigma_{\mu}^2$. Alternative weights have the advantage of decreasing the impact of the brightest, lowest-$\sigma_{\mu}$ points in the convergence procedure, resulting in a solution which is closer to the true value.
\end{multicols}

To assess whether the obtained results emerge simply as a consequence of testing purely theoretical models, the exercise was repeated by inducing perturbations to the SBPs so that:

\vspace*{-0.5cm}
\begin{equation}
\mu(R) = g(\mu(R)^{\rm mod},\sigma_{\mu}),
\label{eqB2}
\end{equation}

where $g(\mu(R)^{\rm mod},\sigma_{\mu})$ is a random deviate from a Gaussian distribution with mean $\mu(R)$ and standard deviation $\sigma_{\mu}$.

\captionsetup{width=18cm}
\begin{figure}[h]
\makebox[\textwidth][c]{\includegraphics[width=1.0\linewidth]{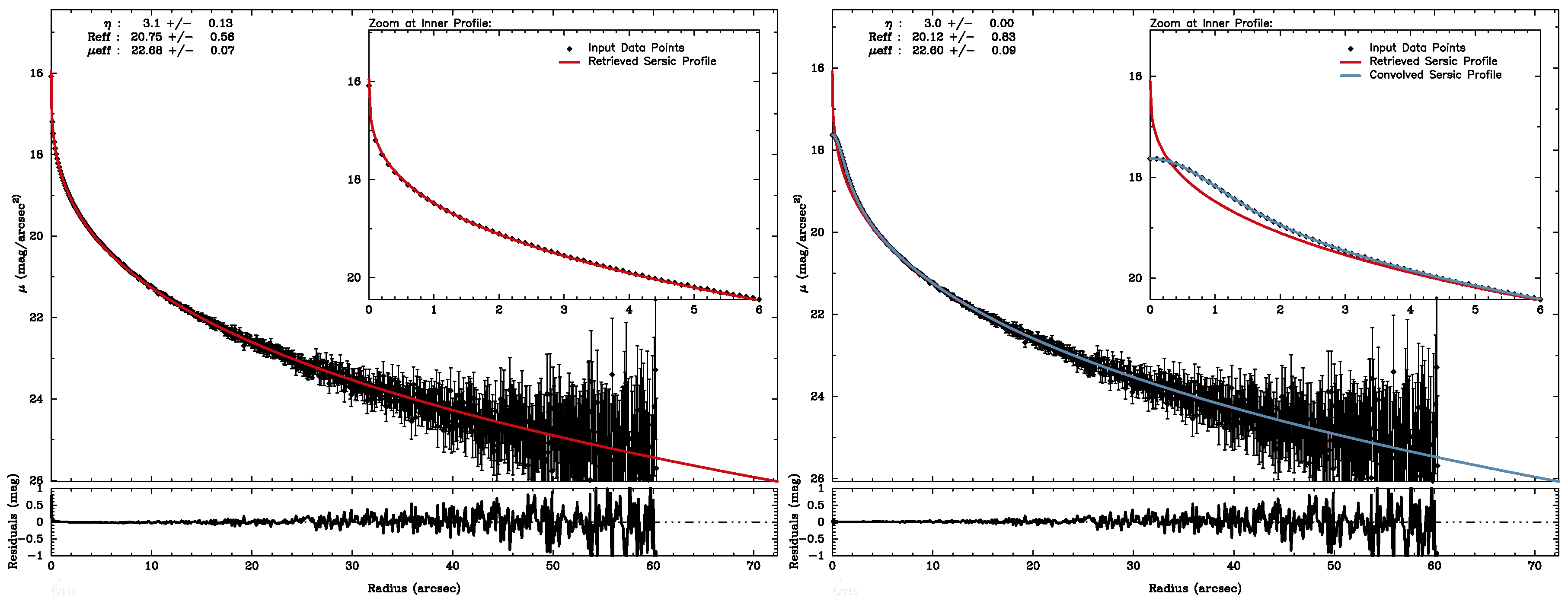}}
\caption[]{
\brem{pA} (left panel) and \brem{pB} (right panel) after random perturbation out of a Gaussian distribution with mean $\mu(R)$ and standard deviation $\sigma_{\mu}$.}\label{Fig_AppB_1_n} 
\end{figure}

\smallskip

\begin{multicols}{2}

\begin{center}
\begin{tabular}{ c | p{0.3cm} c c | p{0.3cm} c c c}
 $w$ & \footnotesize $\eta_3$: & $\rm R_{eff}$ & $\mu_{\rm eff}$ & \footnotesize \ifit: & $\eta$ & $\rm R_{eff}$ & $\mu_{\rm eff}$ \\ 
 no weight & & 20.8 & 22.7 & & 3.3 & 21.4 & 22.8 \\ 
 $1/\sqrt{\sigma_{\mu}}$ & & 20.2 & 22.6 & & 3.1 & 21.1 & 22.7 \\ 
 $1/\sigma_{\mu}$ & & 20.0 & 22.6 & & 3.1 & 20.7 & 22.7 \\ 
 $1/\sigma_{\mu}^2$ & & 20.1 & 22.6 & & 3.0 & 20.1 & 22.6 \\ 
\end{tabular}
\end{center}

\columnbreak

\begin{center}
\begin{tabular}{ c | p{0.3cm} c c | p{0.3cm} c c c}
 $w$ & \footnotesize $\eta_3$: & $\rm R_{eff}$ & $\mu_{\rm eff}$ & \footnotesize \ifit: & $\eta$ & $\rm R_{eff}$ & $\mu_{\rm eff}$ \\ 
 no weight & & 21.1 & 22.7 & & 3.0 & 21.1 & 22.7 \\ 
 $1/\sqrt{\sigma_{\mu}}$ & & 20.1 & 22.6 & & 3.0 & 20.1 & 22.6 \\ 
 $1/\sigma_{\mu}$ & & 19.6 & 22.6 & & 3.0 & 19.6 & 22.6 \\ 
 $1/\sigma_{\mu}^2$ & & 19.1 & 22.5 & & 3.0 & 19.1 & 22.5 \\ 
\end{tabular}
\end{center}

\end{multicols}

\clearpage

\begin{multicols}{2}

\captionsetup{width=9cm}
\begin{Figure1}
\includegraphics[width=1.0\linewidth]{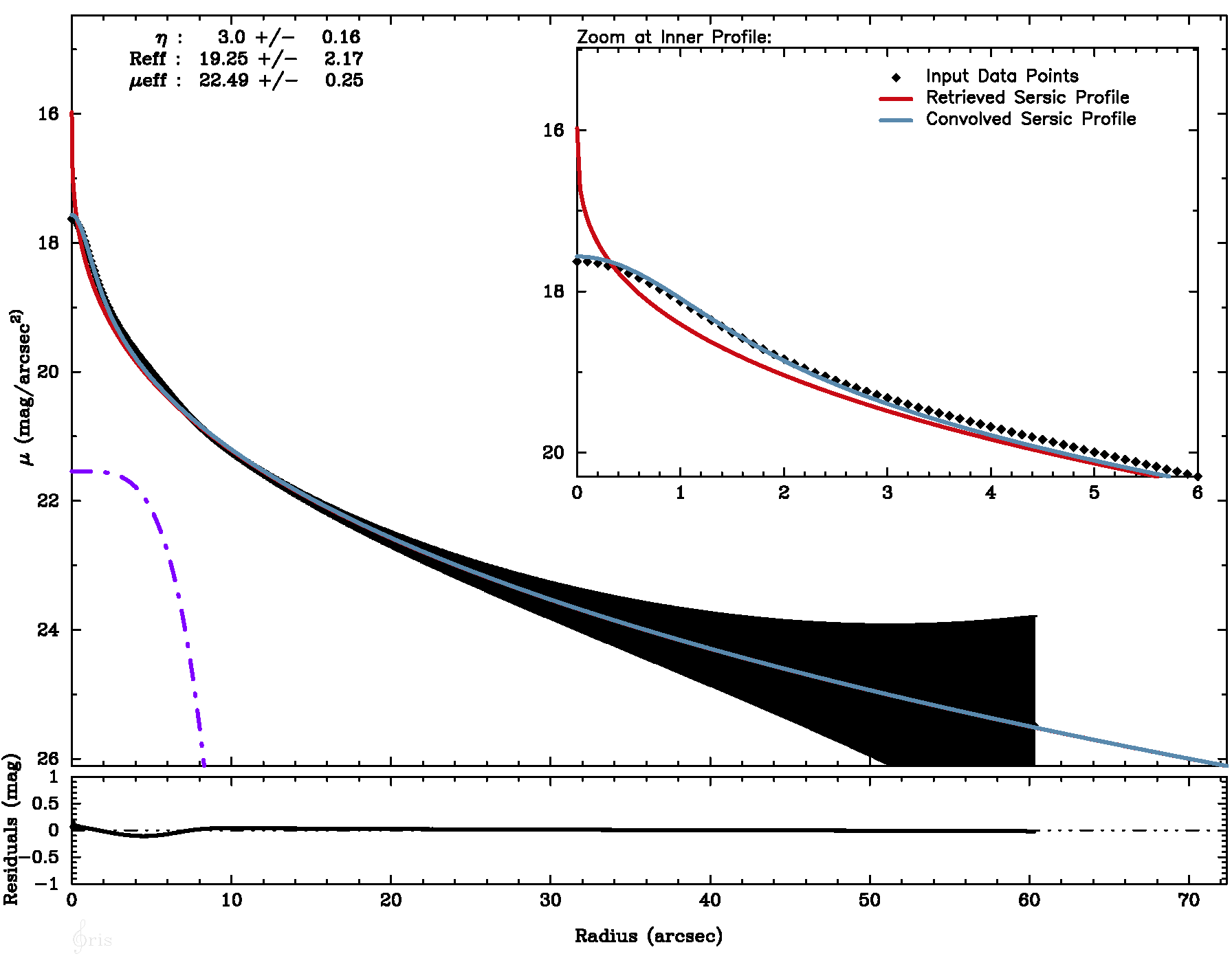}
\captionof{figure}{theoretical \SL\ profile \brem{pB} plus a subtle additional component overplotted in purple, modeled by Eq.~\ref{eq1} with $\eta = 0.25$, $\mu_0 = 21.54$, $\alpha = 5.79$ (\SL\ profile \brem{pC}), with the \ifit\ solution overplotted.}\label{Fig_AppB_2} 
\end{Figure1}  

\columnbreak

3) Adding a subtle luminosity component modeled by Eq.~\ref{eq1} with $\eta = 0.25$, $\mu_0 = 21.54$, $\alpha = 5.79$ to the previous convolved \SL\ profile yields a luminosity increase by $\sim$4\% (Fig.~\ref{Fig_AppB_2}) and the following solutions: 

\begin{center}
\begin{tabular}{ c | p{0.3cm} c c | p{0.3cm} c c c}
 $w$ & \footnotesize $\eta_3$: & $\rm R_{eff}$ & $\mu_{\rm eff}$ & \footnotesize \ifit: & $\eta$ & $\rm R_{eff}$ & $\mu_{\rm eff}$ \\ 
 no weight & & 19.2 & 22.5 & & 3.0 & 19.2 & 22.5 \\ 
 $1/\sqrt{\sigma_{\mu}}$ & & 18.6 & 22.4 & & 2.8 & 18.7 & 22.4 \\ 
 $1/\sigma_{\mu}$ & & 17.6 & 22.3 & & 2.8 & 17.5 & 22.3 \\ 
 $1/\sigma_{\mu}^2$ & & 15.1 & 22.0 & & 2.6 & 14.4 & 21.9 \\ 
\end{tabular}
\end{center}

The results obtained by adopting standard weights are the ones that farthest depart from the correct solution. This occurs because, as previously mentioned, by using standard WLS to fit an \iSP\ that deviates from the \SL\ at low radii such as the \iSP\ in Fig.~\ref{Fig_AppB_2}, low-$\sigma_{\mu}$, bright SBP points which hold the imprints of the extra luminosity component and thus mostly deviate from the \SL\ will dictate the fit. From the table above it is also apparent that alternative weights recover far better the characteristics of the photometrically dominant, underlying \SL\ component.
\end{multicols}

As for the perturbed counterpart:

\begin{multicols}{2}

\captionsetup{width=9cm}
\begin{Figure1}
\includegraphics[width=1.0\linewidth]{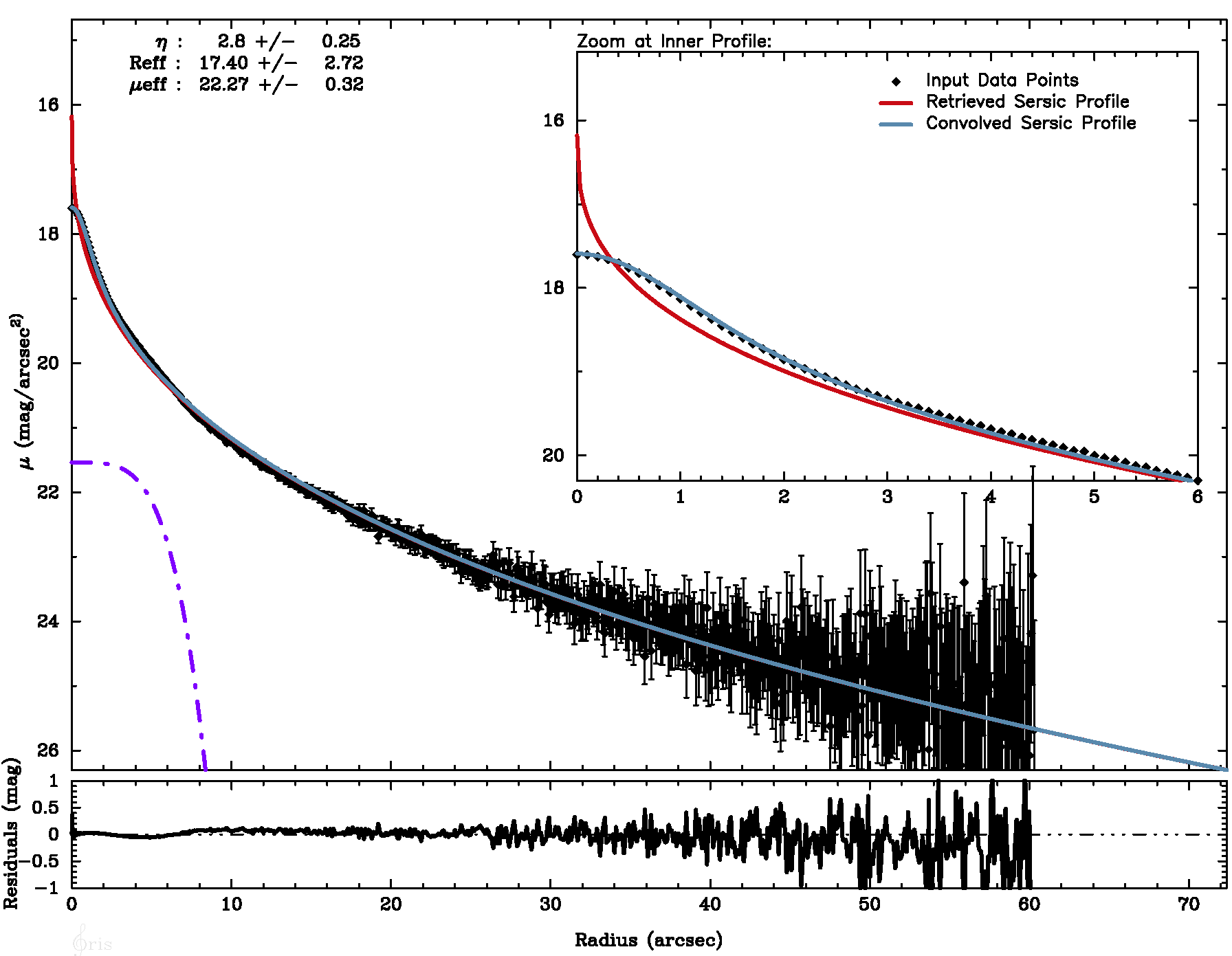}
\captionof{figure}{\brem{pC} after random Gaussian perturbation with mean $\mu(R)$ and standard deviation $\sigma_{\mu}$.}\label{Fig_AppB_2} 
\end{Figure1}  

\columnbreak

\begin{center}
\begin{tabular}{ c | p{0.3cm} c c | p{0.3cm} c c c}
 $w$ & \footnotesize $\eta_3$: & $\rm R_{eff}$ & $\mu_{\rm eff}$ & \footnotesize \ifit: & $\eta$ & $\rm R_{eff}$ & $\mu_{\rm eff}$ \\ 
 no weight & & 20.4 & 22.6 & & 3.1 & 20.3 & 22.6 \\ 
 $1/\sqrt{\sigma_{\mu}}$ & & 18.9 & 22.4 & & 2.9 & 18.9 & 22.4 \\ 
 $1/\sigma_{\mu}$ & & 17.5 & 22.3 & & 2.8 & 17.4 & 22.3 \\ 
 $1/\sigma_{\mu}^2$ & & 14.9 & 22.0 & & 2.5 & 14.0 & 21.9 \\ 
\end{tabular}
\end{center}

\end{multicols}

\clearpage

\captionsetup{width=18cm}
\begin{figure}[h]
\makebox[\textwidth][c]{\includegraphics[width=1.0\linewidth]{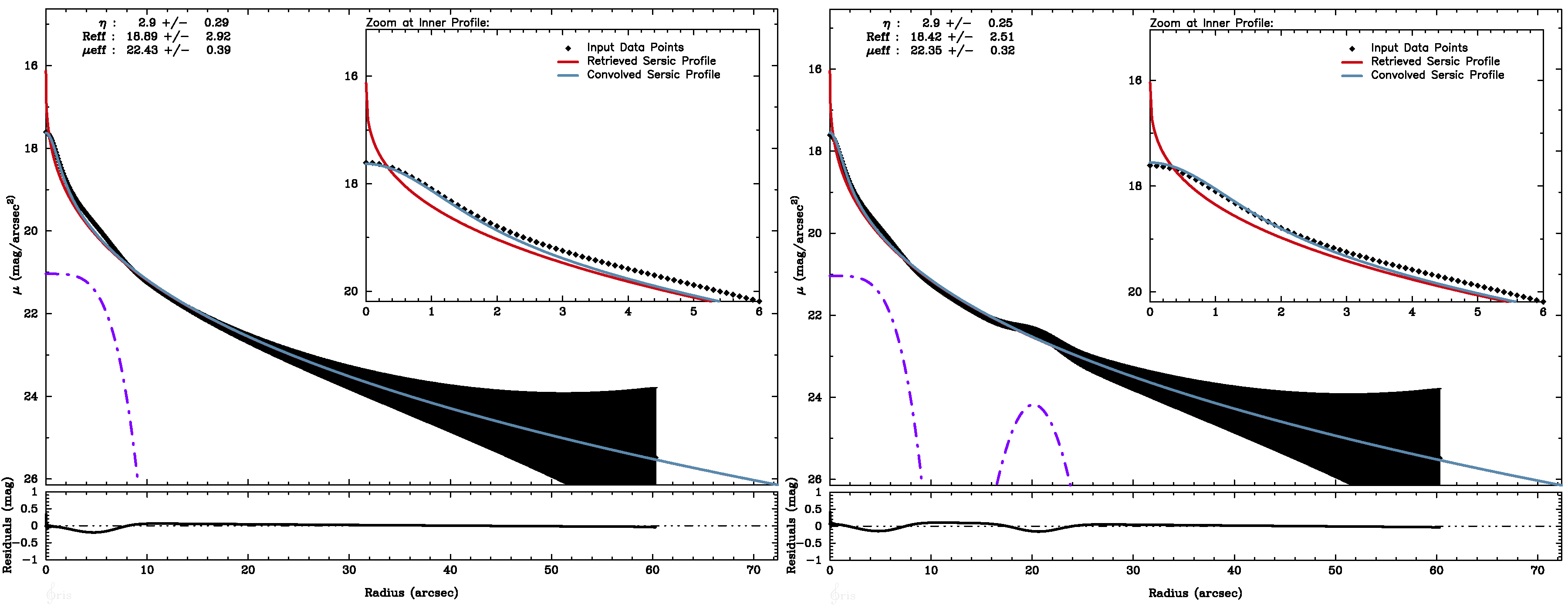}}
\caption[]{\brem{Left panel:} theoretical \SL\ profile \brem{pB} plus an additional component modeled by Eq.~\ref{eq1} with $\eta = 0.25$,  $\mu_0 = 21.04$, $\alpha = 6.20$ (\SL\ profile \brem{pD}) and respective \ifit\ fit. \brem{Right panel:} \brem{pD} plus an extra luminosity excess in the faint end of the \SL\ modeled by the polynomial equation $\mu(R) = 0.148\cdot R^2 - 5.960\cdot R + 84.180$, (\SL\ profile \brem{pE}) with the \ifit\ solution overplotted. Overplotted dashed-dot purple lines display the additional components.}\label{Fig_AppB_4} 
\end{figure} 

\begin{multicols}{2}
4) Adding an additional component given by Eq.~\ref{eq1} with $\eta = 0.25$,  $\mu_0 = 21.04$, $\alpha = 6.20$ to \brem{pB}, enhancing the total luminosity by $\sim$6\%, (profile \brem{pD} at the left-hand side of Fig.~\ref{Fig_AppB_4}) yields the fitting solutions:

\begin{center}
\begin{tabular}{ c | p{0.3cm} c c | p{0.3cm} c c c}
 $w$ & \footnotesize $\eta_3$: & $\rm R_{eff}$ & $\mu_{\rm eff}$ & \footnotesize \ifit: & $\eta$ & $\rm R_{eff}$ & $\mu_{\rm eff}$ \\ 
 no weight & & 18.7 & 22.4 & & 2.9 & 18.9 & 22.4\\ 
 $1/\sqrt{\sigma_{\mu}}$ & & 17.6 & 22.3 & & 2.9 & 17.7 & 22.3\\ 
 $1/\sigma_{\mu}$ & & 16.1 & 22.1 & & 2.6 & 16.1 & 22.1\\ 
 $1/\sigma_{\mu}^2$ & & 12.8 & 21.7 & & 2.3 & 12.2 & 21.5\\ 
\end{tabular}
\end{center}

documenting a similar behavior as the one discussed for the \iSP\ profile \brem{pC}.

\columnbreak

5) Likewise, profile \brem{pE} (right-hand side of Fig.~\ref{Fig_AppB_4}) simulates a wiggle at intermediate radii of profile \brem{pC} 
through addition of a component modeled by the polynomial equation $\mu(R) = 0.148\cdot R^2 - 5.960\cdot R + 84.180$, enhancing the total luminosity by $\sim$9\%:

\begin{center}
\begin{tabular}{ c | p{0.3cm} c c | p{0.3cm} c c c}
 $w$ & \footnotesize $\eta_3$: & $\rm R_{eff}$ & $\mu_{\rm eff}$ & \footnotesize \ifit: & $\eta$ & $\rm R_{eff}$ & $\mu_{\rm eff}$ \\ 
 no weight & & 18.3 & 22.3 & & 2.9 & 18.4 & 22.4 \\ 
 $1/\sqrt{\sigma_{\mu}}$ & & 17.8 & 22.3 & & 2.7 & 17.9 & 22.3 \\ 
 $1/\sigma_{\mu}$ & & 16.8 & 22.1 & & 2.6 & 16.7 & 22.1 \\ 
 $1/\sigma_{\mu}^2$ & & 13.7 & 21.8 & & 2.3 & 12.9 & 21.6 \\ 
\end{tabular}
\end{center}

It can again be seen that an alternative weight recovers best the characteristics of the dominant \SL\ component, whereas WLF with standard weights underestimates $\eta$ and \reff\ by, respectively, 23\% and 36\%, while they overestimate $\mu_{\rm eff}$ by one mag.
\end{multicols}

\clearpage

Finally:

\captionsetup{width=18cm}
\begin{figure}[h]
\makebox[\textwidth][c]{\includegraphics[width=1.0\linewidth]{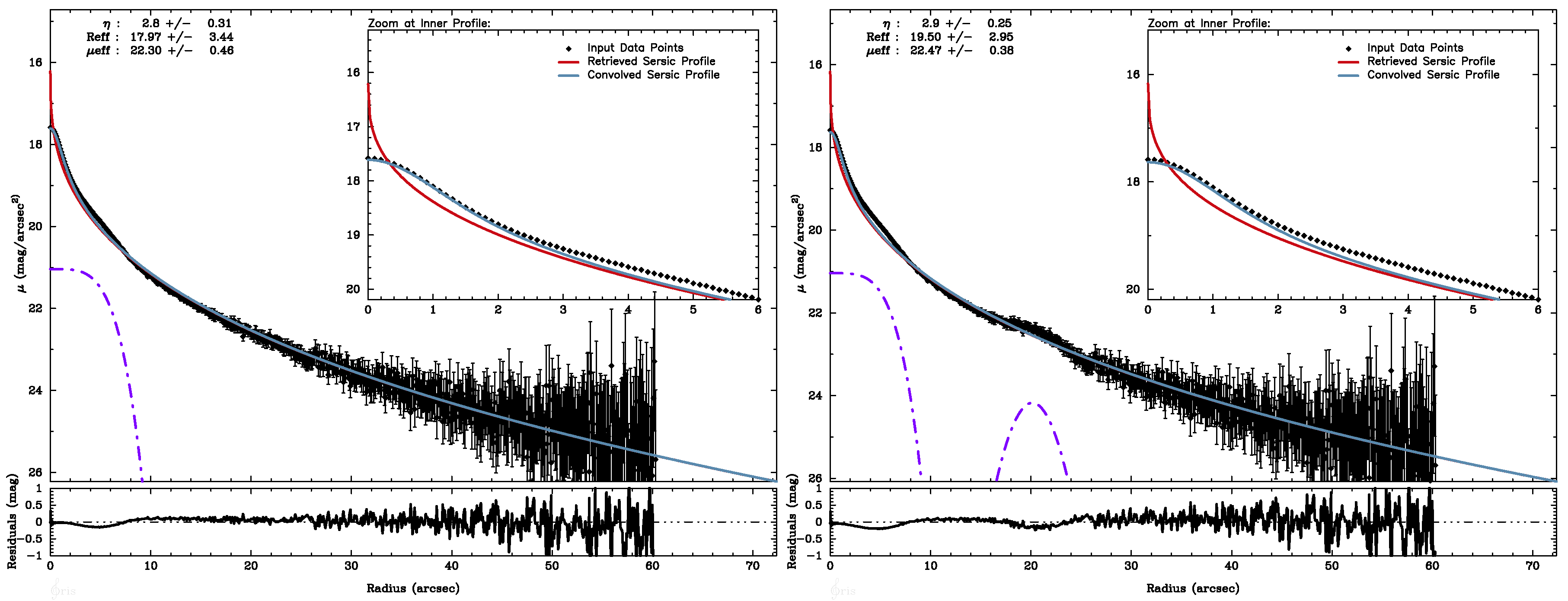}}
\caption[]{
\brem{pD} (left panel) and \brem{pE} (right panel) after random perturbation out of a Gaussian distribution with mean $\mu(R)$ and standard deviation $\sigma_{\mu}$.}\label{Fig_AppB_5} 
\end{figure} 

\begin{multicols}{2}

\begin{center}
\begin{tabular}{ c | p{0.3cm} c c | p{0.3cm} c c c}
 $w$ & \footnotesize $\eta_3$: & $\rm R_{eff}$ & $\mu_{\rm eff}$ & \footnotesize \ifit: & $\eta$ & $\rm R_{eff}$ & $\mu_{\rm eff}$ \\ 
 no weight & & 19.9 & 22.5 & & 2.9 & 19.9 & 22.6 \\ 
 $1/\sqrt{\sigma_{\mu}}$ & & 17.9 & 22.3 & & 2.8 & 18.0 & 22.3 \\ 
 $1/\sigma_{\mu}$ & & 16.0 & 22.1 & & 2.6 & 15.9 & 22.0 \\ 
 $1/\sigma_{\mu}^2$ & & 12.6 & 21.6 & & 2.2 & 11.9 & 21.5 \\ 
\end{tabular}
\end{center}

\columnbreak

\begin{center}
\begin{tabular}{ c | p{0.3cm} c c | p{0.3cm} c c c}
 $w$ & \footnotesize $\eta_3$: & $\rm R_{eff}$ & $\mu_{\rm eff}$ & \footnotesize \ifit: & $\eta$ & $\rm R_{eff}$ & $\mu_{\rm eff}$ \\ 
 no weight & & 19.3 & 22.5 & & 2.9 & 19.5 & 22.5 \\ 
 $1/\sqrt{\sigma_{\mu}}$ & & 18.0 & 22.3 & & 2.7 & 18.1 & 22.3 \\ 
 $1/\sigma_{\mu}$ & & 16.7 & 22.1 & & 2.6 & 16.5 & 22.1 \\ 
 $1/\sigma_{\mu}^2$ & & 13.5 & 22.5 & & 2.3 & 12.7 & 21.6 \\ 
\end{tabular}
\end{center}

\end{multicols}

The same result is observed for a set of diverse \iSP's with or without noise: the standard WLS yield incorrect solutions, in the sense of the largest deviations from the characteristics of the photometrically dominant, underlying \SL\ component. As discussed above, this occurs due to the nature of the $\chi^2$ minimization method, that largely bases the solution on the brightest, lowest-$\sigma_{\mu}$ points of a SBP (or galaxy image). 

Although with these examples we tried to simulate some of the typically observed \iSP's, it is obviously impossible to illustrate and carry out a comparative \SL\ modeling of all possible \iSP's. For this reason, and given that for some \iSP's the standard WLS or an intermediate weight is the one that better recovers the true \SL\ parameters, \ifit\ is designed such as to employ the four weighting methods and subsequently converge to the solution which best matches the observed light growth curve.

\end{appendices}
\end{document}